\documentclass[a4paper,11pt]{article}
\pdfoutput=1 
\usepackage{jcappub}
\usepackage{amsmath}
\usepackage{amssymb}
\usepackage{rotating}
\usepackage[T1]{fontenc} 
\usepackage{subcaption}
\usepackage{float}
\usepackage{graphicx}

\def\({\left(}
\def\){\right)}
\def\[{\left[}
\def\]{\right]}
\def\be{\begin{equation}}
\def\ee{\end{equation}}
\def\beq{\begin{eqnarray}}
\def\eeq{\end{eqnarray}}

\title{Prospects of constraining reionization model parameters using Minkowski tensors and Betti numbers}

\author[a,b]{Akanksha Kapahtia}
\author[b]{Pravabati Chingangbam}
\author[c,d]{Raghunath Ghara}
\author[e,f]{Stephen Appleby}
\author[a]{Tirthankar Roy Choudhury}

\affiliation[a]{National Center for Radio Astrophysics,  TIFR, Ganeshkhind, Pune 411007, India}
\affiliation[b]{Indian Institute of Astrophysics, Koramangala II Block,       
  Bangalore  560 034, India}

\affiliation[c]{Department of Physics, Technion, Haifa 32000, Israel}
\affiliation[d]{Department of Natural Sciences, The Open University of Israel,Ra'anana 4353701, Israel}

\affiliation[e] {Asia-Pacific Center for Theoretical Physics (APCTP), Pohang, Republic of Korea}
\affiliation[f]{Quantum Universe Center, Korea Institute for Advanced Study, 85 Hoegiro, Dongdaemun-gu, Seoul 02455, Korea}
\emailAdd{akapahtia@ncra.tifr.res.in}

\abstract{We explore the possibility of constraining model parameters of the Epoch of Reionization (EoR) from 21cm brightness temperature maps, using a combination of morphological descriptors constructed from the eigenvalues of the Contour Minkowski Tensor (CMT), Betti numbers (count of connected regions $n_{con}$ and holes $n_{hole}$) and the area of structures in the excursion set of the field. We use a three parameter model of EoR simulated using 21\textrm{cmFAST}, namely the ionizing efficiency of sources $\zeta$, the minimum virial temperature $T_{vir}$ required for collapse into a halo and the maximum radius for ionizing radiation described by $R_{mfp}$. We performed a Bayesian analysis to recover model parameters for a mock 21cm image from SKA phase I at a redshift of $z=7.4$ corresponding to a mean neutral hydrogen fraction of $\mathrm{\bar x}_{HI} \simeq 0.5$.  We find that in the absence of noise the average size of structures in the field with $x_{HI} \lesssim 0.5$ is smaller than regions with $x_{HI} \gtrsim 0.5$ and the structures are equally isotropic when  $\mathrm{\bar x}_{HI}=0.5$ . We also find that in order to recover the input model to within $1-\sigma$ accuracy for a mock noisy image at a single frequency channel of $1~\mathrm{MHz}$, for an observation time $t_{obs}<2000~\mathrm{hrs}$, the noisy $\delta T_b$ map needs to be smoothed at a scale $R_s>9.5~\mathrm{Mpc}$. Finally we show that the systematic behaviour of the statistic as ionization progresses, enables us to obtain stringent constraints on $\mathrm{\bar x}_{HI}$ (with a coefficient of variation $\sim 0.05$ as compared to $\sim 0.1-0.2$ for model parameter constraints), thereby making these descriptors a promising statistic for constraining EoR model parameters and the ionization history of the universe. 
}

\begin{document}
\maketitle
\flushbottom

\section{Introduction}
The baryonic component of the universe post recombination was dominated by neutral hydrogen, which clumped into high density regions leading to the formation of first luminous sources in the universe. The emission from these first sources of radiation, ionized the neutral hydrogen in the intervening medium leading to an important phase transition of the universe called the Epoch of Reionization (EoR) \cite{loeb, Dayal:2018hft}. The epoch is characterized by the appearance of ionized regions around these luminous sources, which gradually grow in size and merge until the entire universe is ionized \cite{gnedin}. Therefore, studying the growth and topology of these ionized regions can help infer important properties of these first luminous sources. One promising observational probe of the EoR is the brightness temperature of the redshifted 21cm transition from neutral hydrogen atoms. The 21cm wavelength redshifts to the range of frequencies accessible by future radio telescopes and would serve as a direct probe of EoR \cite{Pritchard}. Statistically it is detected in terms of the sky averaged global signal or in terms of the power spectrum of the brightness temperature fluctuations. The global signal was recently claimed to have been detected by the EDGES experiment \cite{Bowman:2018} while the detection of the 21cm power spectrum is being aimed at by various ongoing interferometeric experiments such as Murchison Widefield Array (MWA), Low-Frequency Array (LOFAR), Giant Metrewave Radio Telescope (GMRT) and Precision Array for Probing the Epoch of Reionization (PAPER) ~\cite{Tingay et al.(2013),van Haarlem,Paciga:2011, Parsons}. The upper limit on the dimensionless 21-cm power spectrum is $\Delta^2_{21} < (73)^2 \mathrm{mK}^2$ at $k=0.075 \mathrm{h~ c~ Mpc}^{-1}$  as found by LOFAR at a redshift of $z \simeq 9.1$ \cite{lofar}. 

 Given the highly non-Gaussian nature of the process of reionization \cite{cooray,Bharadwaj2005,shaw,Banet}, one would need to employ statistical tools which encapsulate information of higher order correlations. For a Gaussian random field the power spectrum contains all the spatial information and no new information is obtained from higher order correlations. Therefore, to probe the EoR one needs to resort to higher order Fourier statistics of the 21cm brightness temperature, such as the bispectrum \cite{Shimabukuro,Watkinson_2017,suman,Giri_bi,hutter2020} or use the phase space statistics \cite{Gorce2019} to infer non-gaussian features of the 21cm signal.

 Complementary to the Fourier space measures are real space topological tools which can be directly employed on 21cm images from upcoming interferometers such as the Square Kilometre Array (SKA)\cite{Maartens et al.(2015)} which would enable direct imaging from the EoR \cite{Mellema} (in addition to better constraints on power spectrum). The advantage of imaging is that it contains both phase and amplitude information of a field. Since information from all orders of correlation are contained in an image, it encapsulates the non-Gaussian information which is otherwise missing in the power spectrum. There are methods which rely on studying the size statistics of ionized bubbles in tomographic images \cite{ giri18, giri19} or based on mathematical morphology \cite{Kakiichi2017,Busch,gazagnes20}. However, these methods rely on first identifying or defining an ionized bubble in a 21cm map which can be difficult in noisy 21cm images \cite{Giri_super}.  There are methods based on percolation theory~\cite{Bag:2018zon,Bag:2018fyr,Furlanetto:2016} and persistent topology \cite{Elbers} which have been used to theoretically study the topological phases of ionized bubbles during EoR.
 There are also methods based on Scalar Minkowski Functionals (SMFs) ~\cite{Lee:2007dt,Friedrich:2010nq,Ahn:2010hg,Wang:2015dna,Gleser:2006su,Yoshiura:2015,Chen} for ionization field or the 21cm brightness temperature field during EoR. These methods can be used to study the topology of an entire field and its systematic variation with field threshold, which can give insights useful for discriminating different EoR scenarios.
 
 The tensorial generalization of SMFs called Minkowski Tensors provide richer morphological information regarding the shape and relative orientation of structures~\cite{Schroder2D:2009} and were recently introduced for cosmological datasets~\cite{Vidhya:2016,Chingangbam:2017,Appleby:2017uvb,Goyal}. In the context of reionization, one of the Minkowski tensors in 2D called \textit{Contour Minkowski Tensor} (CMT) was used along with Betti numbers to capture the ionization history during EoR in terms of lengthscale and timescale information encapsulated in these descriptors~\cite{Kapahtia:2017qrg}. In a subsequent study in \cite{Kapahtia2019} the authors showed that the said combination of morphological descriptors serve as a useful tool to discriminate different EoR scenarios. These studies were based on ideal noiseless realizations of the 21cm brightness temperature field. 
  
In this paper we perform a more quantitative study by employing CMT in combination with Betti numbers to explore the prospects of constraining a simple three parameter model of EoR using mock 21cm images from SKA I low\footnote{http://skatelescope.org}. In addition to using the scale information contained in CMT, here we also use the area of structures in the excursion sets of the field. For this study, we simulate 21cm brightness temperature maps at a redshift of $\mathrm{z=7.4}$ using 21cm\textrm{FAST} \cite{Mesinger:2010ne} and perform a Bayesian inference using COSMOMC \cite{cosmomc} as a generic sampler. A detailed study of detecting ionized sources from 21cm images in the presence of system noise and foreground has been performed in  \cite{ghara2017}. In this work, we will follow a similar method for generating system noise and perform our inference for a model of EoR described by the ionizing efficieny of sources ($\zeta$), the minimum Virial temperature for collapse into a halo ($T_{vir}$) and the maximum scale of ionizing radiation ($R_{mfp}$). The study is performed by assuming complete foreground removal from the mock 21cm images.  

This paper is organized as follows. In section~\ref{sec:2} we give a brief review of the morphological descriptors used for this study. In section~\ref{sec:3.1} we give a brief description of 21cmFAST used for simulating the 21-cm differential brightness temperature field, (or $\delta T_b$) for this study. Then we discuss how the mock observation from SKA-low is constructed for this study in section~\ref{mock}. In section~\ref{parameter} we discuss the physics behind the behaviour of the morphological statistics in parameter space, followed by a discussion of the impact of noise and smoothing on the statistics in section~\ref{bias}. In section~\ref{Bayes} we describe our methodology of the Bayesian analysis for obtaining the results described in section~\ref{results}. We end with a summary of our results and conclusion in section ~\ref{conc}.

\section{Morphological Descriptors}
\label{sec:2}
For any field $f(x)$ the set of all field values greater than or equal to a certain field threshold, $f=u_{th}$ is called an {\it{excursion set}}. The boundary curves of these excursion sets in two dimensions enclose either a connected region (regions formed by bounded curves enclosing a set of values greater than or equal to $u_{th}$) or a hole (regions formed by bounded curves enclosing a set of values less than $u_{th}$). The number of connected regions, $n_{con}$ and holes, $n_{hole}$ at the threshold $u_{th}$ are called {\it{Betti numbers}}. Minkowski functionals describe the morphology and topology of excursion set regions for a given random field as a function of field threshold $u_{th}$. The morphology and number of these excursion set regions changes as $u_{th}$ is varied.
Minkowski Tensors~(MTs) are tensor generalization of the Scalar Minkowski Functionals.  We will focus on the translation invariant symmetric rank two tensor, which we refer to as the {\em contour} MT (CMT), defined for a single boundary curve $C$  as \cite{Schroder2D:2009}:
\begin{equation}
\mathcal{W}_{1}=\int_C \hat{T}\otimes \hat{T} ~{\rm d}s,
\label{eqn:W1}
\end{equation}
where $\hat{T}$ is the unit tangent vector at every point on the curve, $\otimes$ denotes the symmetric tensor product given by
\begin{equation}
\left(\hat{T}\otimes \hat{T}\right)_{ij} = \frac12\left( \hat{T}_i\hat{T}_j + \hat{T}_j\hat{T}_i\right),
\end{equation}
and ${\rm d}s$ is the infinitesimal arc length. In \cite{Schroder2D:2009,Chingangbam:2017,Appleby:2017uvb} $\mathcal{W}_{1}$ is referred to as ${W}_2^{1,1}$.  It was shown in \cite{Schroder2D:2009,Chingangbam:2017} that it is the only linearly independent translation invariant Minkowski tensor that contains extra information over its scalar counterpart. The extra information contained in  $\mathcal{W}_{1}$ is that of anisotropy. Any anisotropy in the boundary curve will manifest as an inequality between the eigenvalues of the matrix ${\cal W}_{1}$. We define the eigenvalues in ascending order, $\lambda_1<\lambda_2$ and define the {\em shape anisotropy parameter} as $\beta\equiv\lambda_1/\lambda_2$. Hence for a generic curve, $\beta$ will have values between 0 and 1. The CMT also gives an estimate of length of the enclosing curve, since $\mathbf{Tr}\left(\mathcal{W}_{1}\right)$ is the perimeter of the curve and related to the second scalar MF i.e. the total contour length denoted by $W_1$.  If $\lambda\equiv (\lambda_1+ \lambda_2)$ denotes the perimeter of the closed curve and is equated to the circumference of a circle i.e. $2\pi r$, we define $r$ to be an \textit{effective radius}:
\begin{equation}
r\equiv\lambda/2\pi.
\label{circle}
\end{equation}
This number gives the largest possible average radius of the enclosing contour. 

We follow the method in \cite{Appleby:2017uvb} based on \textit{Marching square algorithm} and \textit{Friend of Friends} algorithm to identify contours surrounding connected regions and holes in the excursion set.
Since our field is dicretized into pixels, we shall use ${\cal{W}}_1$ for a polygon \cite{Schroder2D:2009} given by:
\begin{eqnarray}
({\cal{W}}_1)_{ab}= \sum_{a,b} |\vec{e_{ab}}|^{-1} \vec{e}_{ab} \otimes \vec{e}_{ab},
\end{eqnarray}
where $|\vec{e}_{ab}|$ is the length of a two dimensional vector describing a discretized segment of the boundary curve between two vertices \textit{a} and \textit{b}.

At each field threshold $u_{th}$ and at a given redshift $z$, we denote the number of distinct curves by $n_{x}(u_{th},z)$, where the suffix $\rm x=con $ or $\rm x=hole$ refers to the boundaries of connected regions or holes respectively. 
Thus, we define statistics averaged over all curves at a fixed $u_{th}$ at a given redshift $z$ as:
\begin{eqnarray}
 {\overline{r}}_{\rm x}(u_{th},z) &\equiv& \frac{\sum_{j=1}^{n_{\rm x}(u_{th},z)} r_{\rm x}(j,z)}{n_{\rm x}(u_{th},z)}, \label{eqn:rx_nu}\\
  \quad {\overline\beta}_{\rm x}(u_{th},z) &\equiv&  \frac{\sum_{j=1}^{n_{\rm x}(u_{th},z)} \beta_{\rm x}(j,z)}{n_{\rm x}(u_{th},z)}.
  \label{eqn:betax_nu}
\end{eqnarray}
${\overline{r}}_{\rm x}(u_{th},z)$ and $  {\overline{\beta}}_{\rm x}(u_{th},z)$ are the mean values at each $u_{th}$ for a redshift $z$.

Next, we define the integrated quantity, $N_{\rm x}$, as,
\begin{equation}
  N_{\rm x} \equiv \int_{u_{\rm low}}^{u_{\rm high}} {\rm d}u_{th} \,n_{\rm x}(u_{th},z),
  \label{eqn:betti}
\end{equation}

Finally, we define the threshold integrated \textit{characteristic} scale and shape anisotropy parameter at a given redshift $z$, using ${\overline{r}}_{\rm x}(u_{th},z)$ and $  {\overline{\beta}}_{\rm x}(u_{th},z)$, as follows:
\begin{eqnarray}
r^{\rm ch}_{\rm x} (z) &\equiv& \frac{\int_{u_{\rm low}}^{u_{\rm high}} {\rm d}u_{th} \,n_{\rm x}(u_{th},z) {\bar{r}}_{\rm x}(u_{th},z)}{N_{\rm x}}, \label{eqn:rch}\\
\beta^{\rm ch}_{\rm x}(z) &\equiv& \frac{\int_{u_{\rm low}}^{u_{\rm high}} {\rm d}u_{th} \,n_{\rm x}(u_{th},z) {\bar{\beta}}_{\rm x}(u_{th},z)}{N_{\rm x}}.
\label{eqn:betach}
\end{eqnarray}

\begin{figure}
	\centering
	\includegraphics[height=6cm,width=7cm]{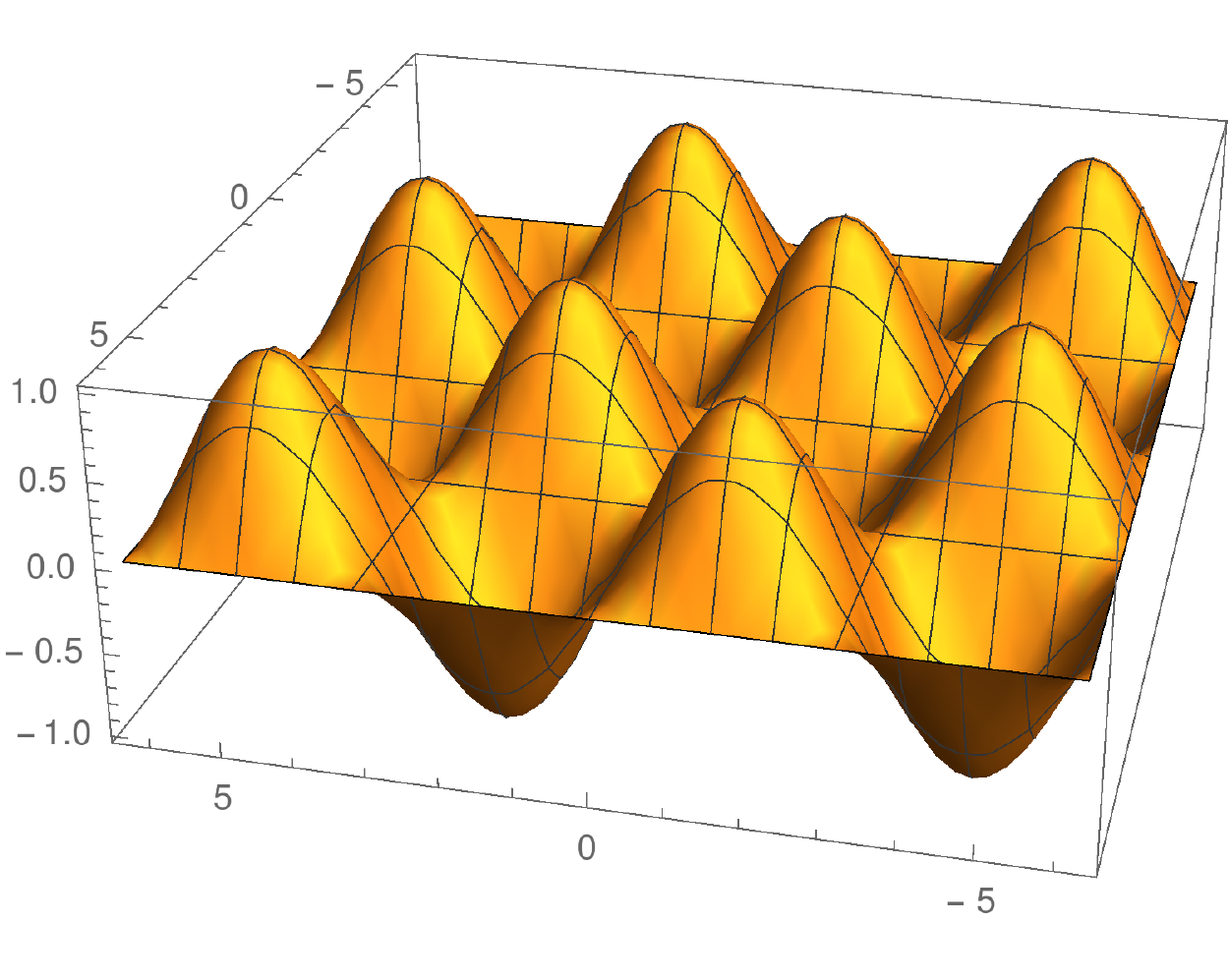}\\
	\includegraphics[height=3.8cm,width=3.5cm]{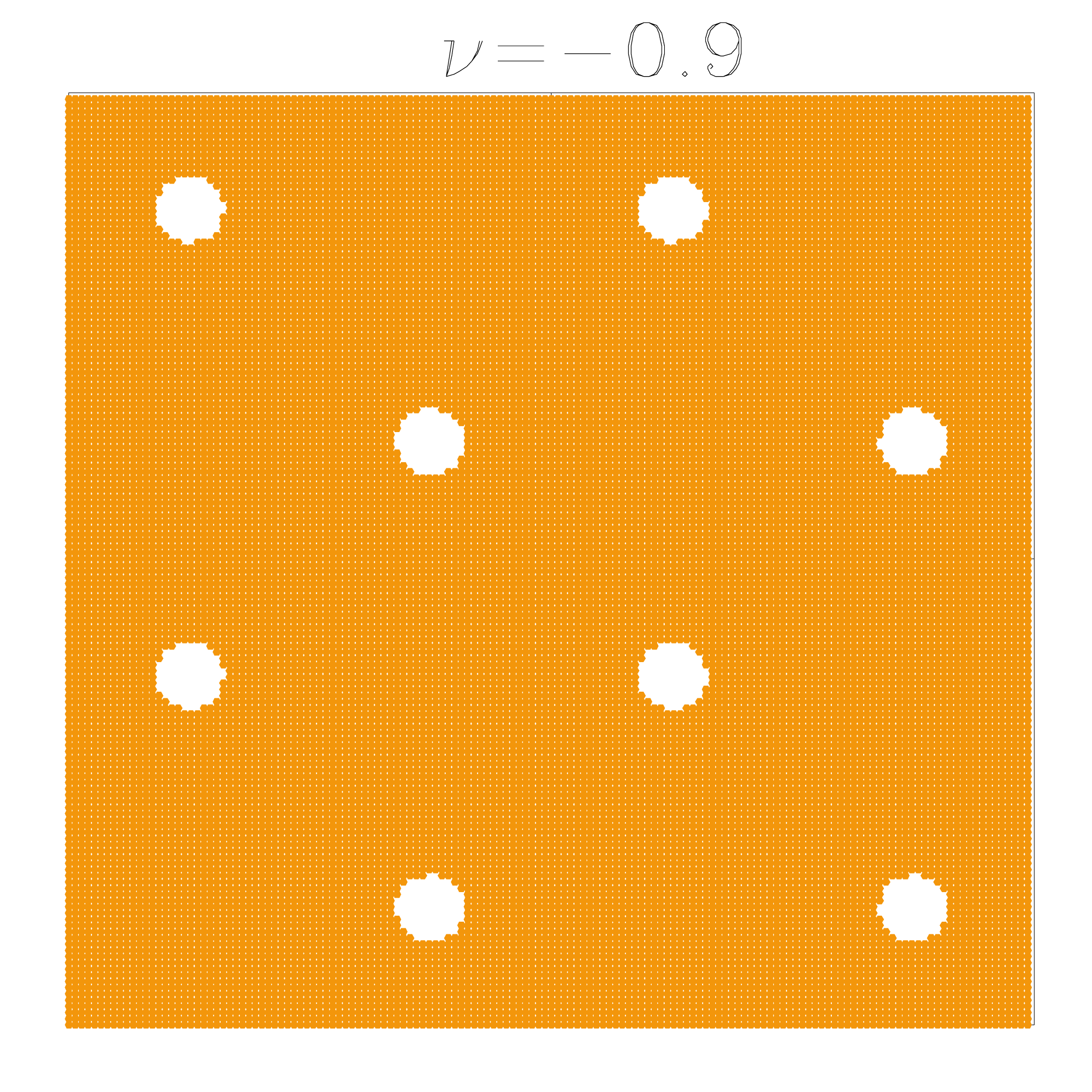}
	\includegraphics[height=3.8cm,width=3.5cm]{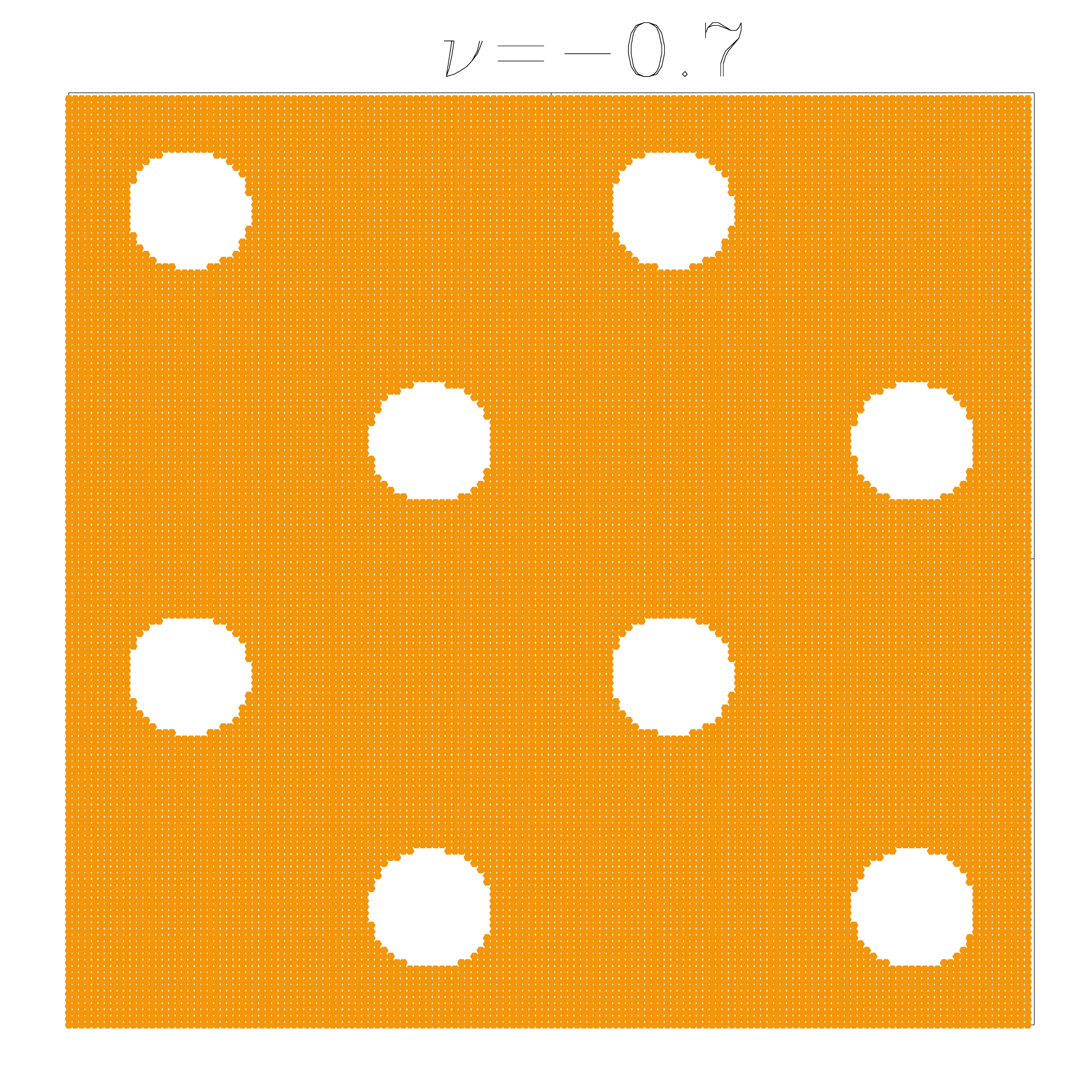}
	\includegraphics[height=3.8cm,width=3.5cm]{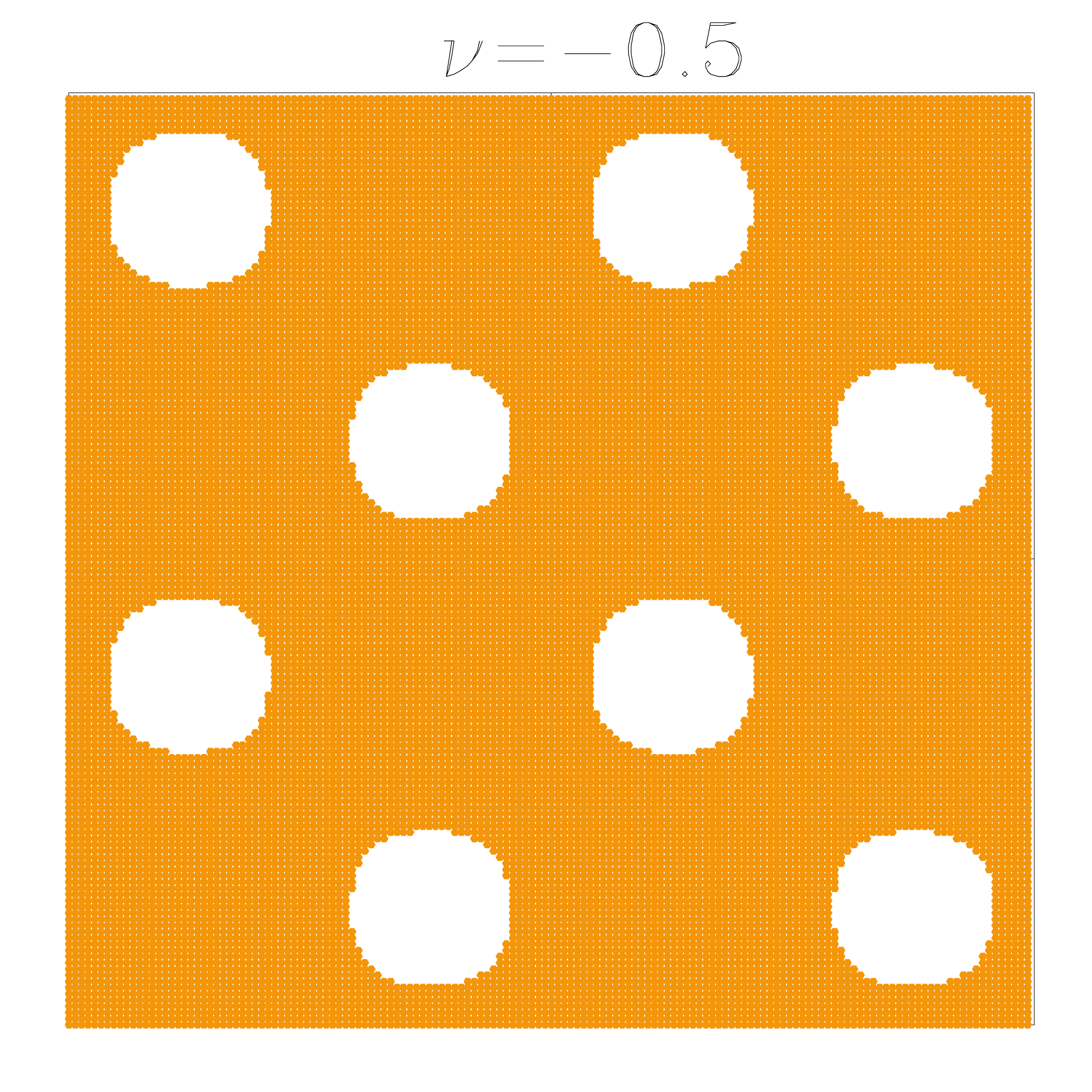}
	\includegraphics[height=3.8cm,width=3.5cm]{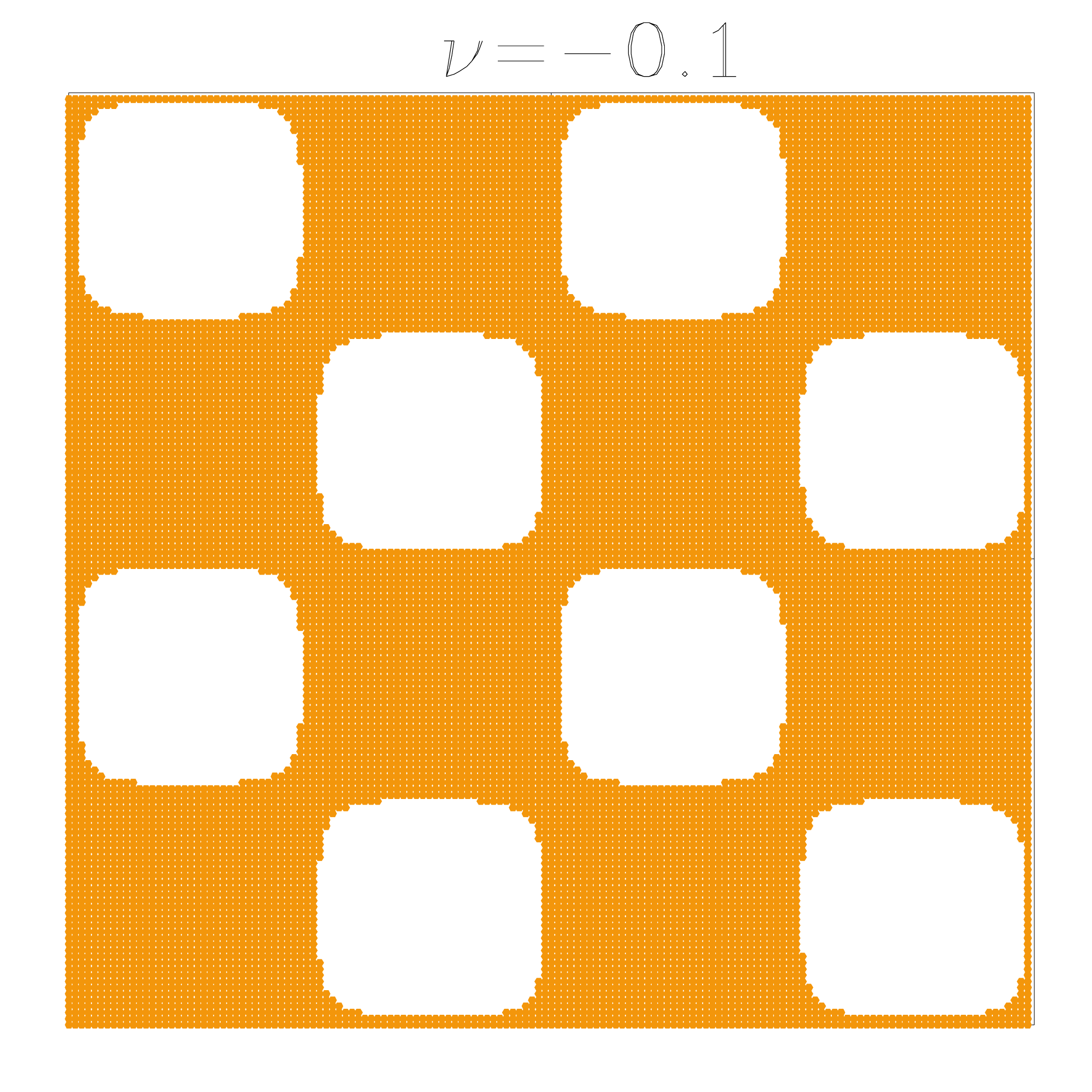}\\
	\includegraphics[height=3.8cm,width=3.5cm]{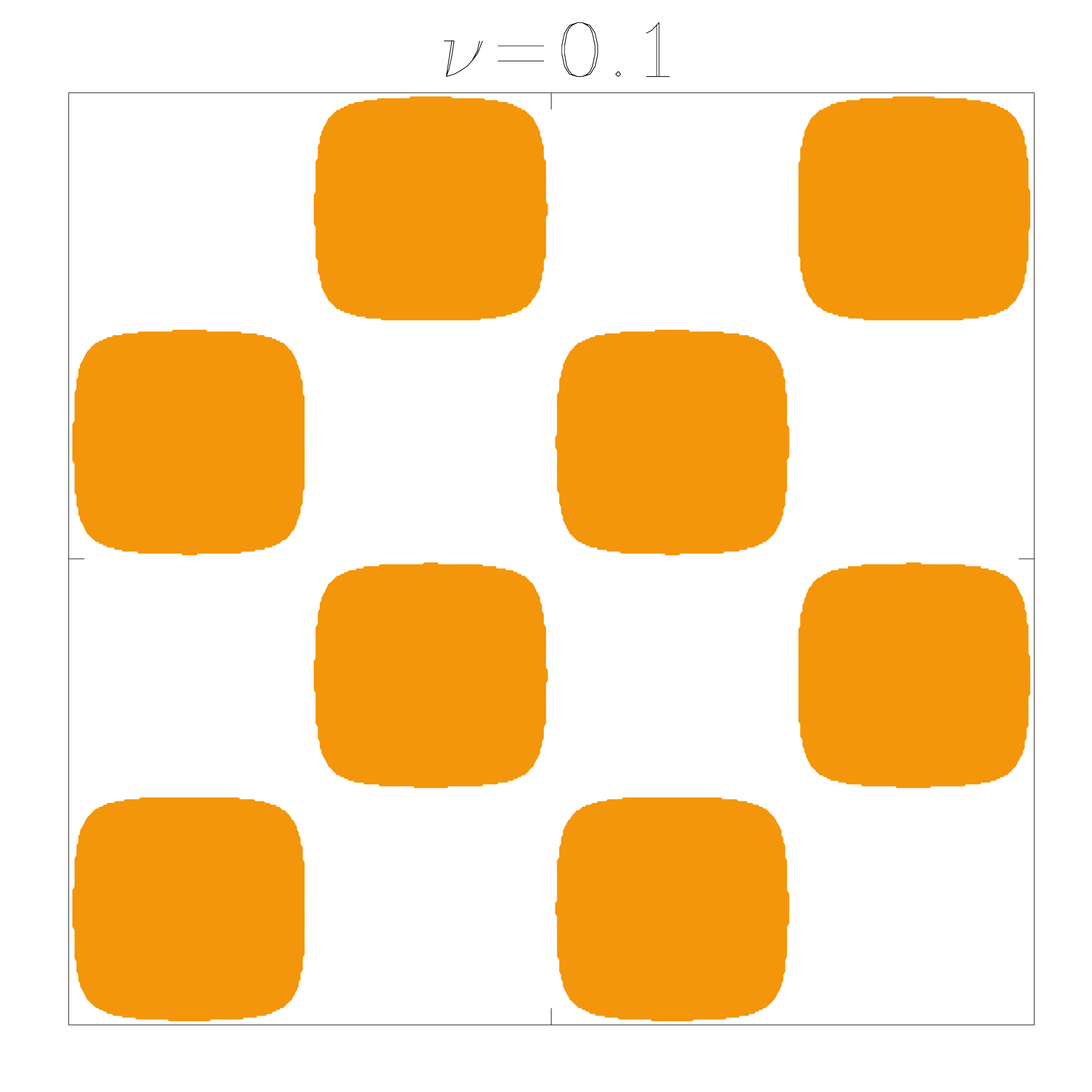}
	\includegraphics[height=3.8cm,width=3.5cm]{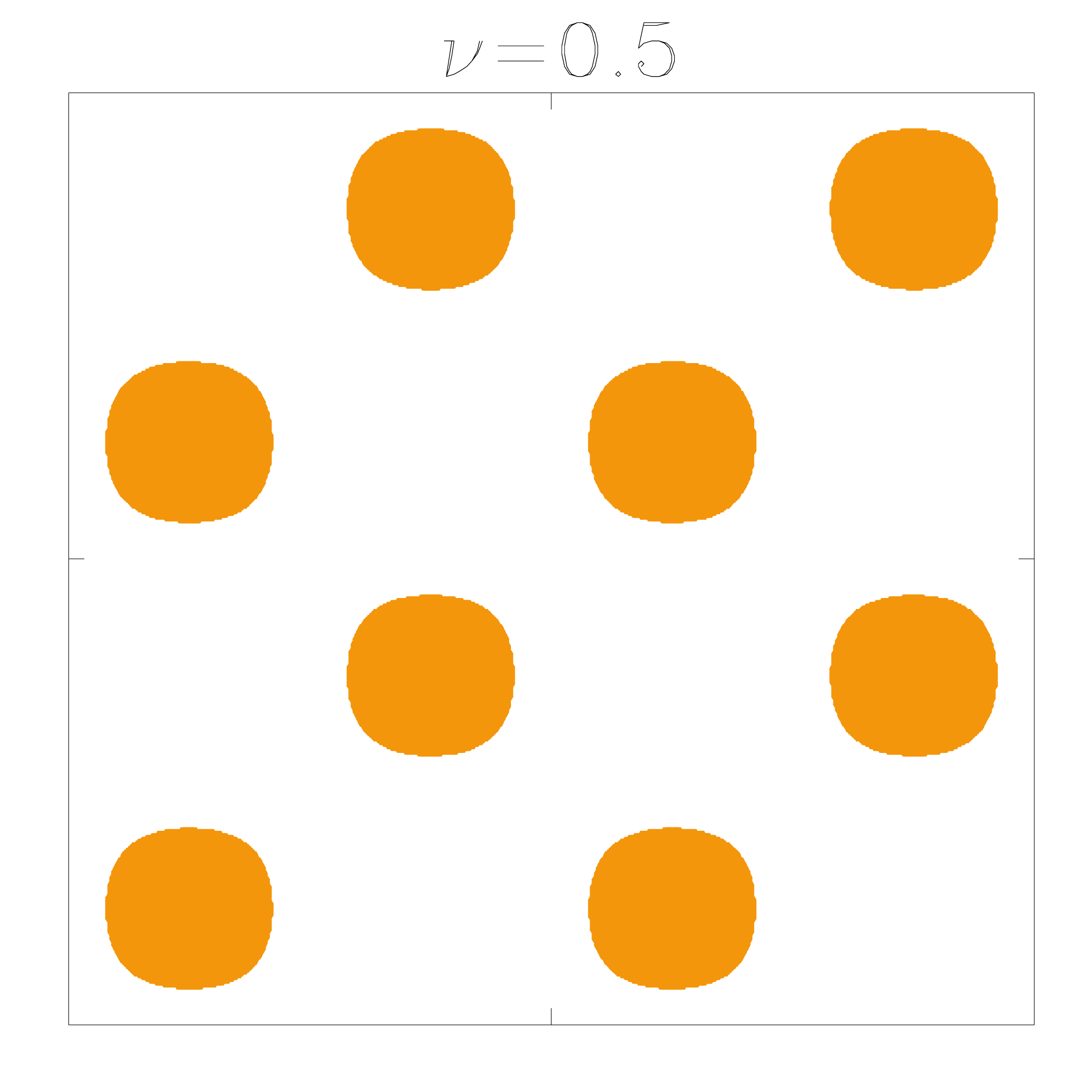}
	\includegraphics[height=3.8cm,width=3.5cm]{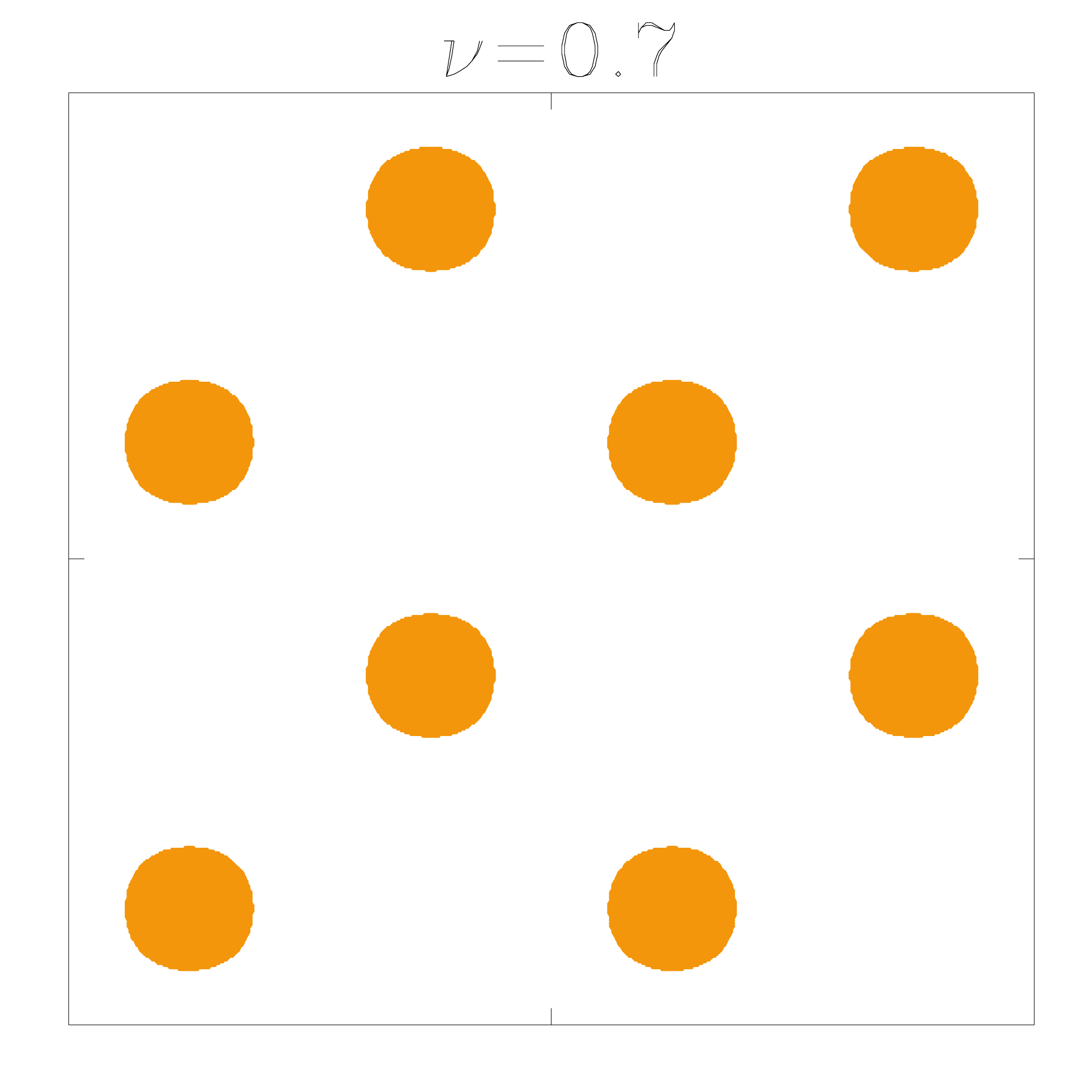}
	\includegraphics[height=3.8cm,width=3.5cm]{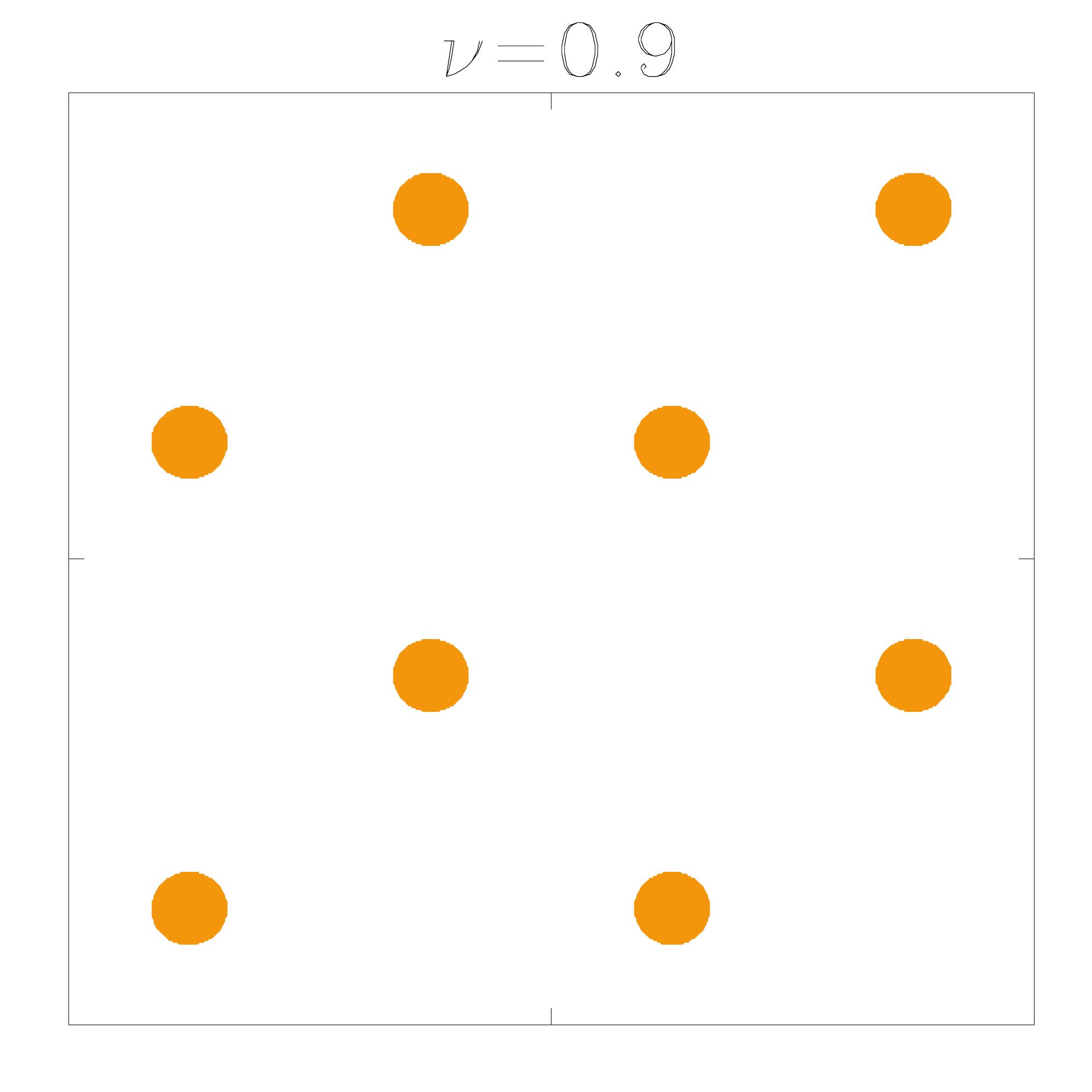}
	\caption{{\em Top:} The function $u(x,y)=\sin(x) \sin(y)$ as a simplistic representation of a random field. {\em Middle}: Excursion sets of $u(x,y)$ at different negative field threshold values which consist of one connected region (orange) and 8 holes (white). The size of the connected region decreases with $\nu \rightarrow 0$, while that of the holes increase. {\em Bottom}: Excursion sets at different positive threshold values which consist of 8 connected regions and 0 holes. The size of the connected regions decrease as $\nu \rightarrow 1$.}
	\label{fig:toy_eset}
\end{figure}
In order to obtain an intuitive understanding of the statistical measures defined above, let us consider a field defined by a 2 dimensional function, $u(x,y)=\sin(x) \sin(y)$. The field values range from -1 to 1 and $\nu = u_{th}$ denote field threshold values. Fig.~\ref{fig:toy_eset} shows $u(x,y)$ (top) and the excursion sets for negative (middle) and positive (bottom) field threshold values. Let us consider holes at negative threshold values and explicitly calculate the statistics for this simple field, as follows:
\begin{enumerate}
	\item We can see from the figure that $n_{hole}=8$ at each negative $\nu$. Suppose we choose the range of $\nu$ integration to be -0.9 to -0.1. Then from eq. \ref{eqn:betti}, we get $N_{hole}=8\times0.8 = 6.4$. 
	\item We see that at each $\nu$ all holes have identical size. Let us denote it by $r_{hole}(\nu)$. So from eq.~\ref{eqn:rx_nu} we get $\bar{r}_{hole}(\nu)=r_{hole}(\nu)$. Then again using integration range -0.9 to -0.1 from eq.~\ref{eqn:rch} we get   $r^{ch}_{hole}(\nu)=r_{hole}(\nu)$. As $\nu\rightarrow 0$ we see that $r_{hole}(\nu)$ decreases (for this toy model). 
	\item Next, we see that at each $\nu$ all holes have identical shape whose corresponding $\beta$  we  denote by $\beta_{hole}(\nu)$. (For this simple example  $\beta_{hole}(\nu)$ is actually independent of $\nu$ but we write the argument for the sake of generality). Then, using the same integration range as above, we obtain from eq.~\ref{eqn:betach} that   $\beta^{ch}_{hole}(\nu)=\beta_{hole}(\nu)$.
\end{enumerate}
It is straightforward to carry out similar calculation for connected regions on the positive threshold side. For general smooth random fields an excursion set at some arbitrary threshold will consist of some number of holes and connected regions, of arbitraty sizes and shapes. So, the analytic calculation of the statistics will in general not be feasible.

It is to be noted that in \cite{Kapahtia:2017qrg} and \cite{Kapahtia2019}, the lower and upper limits of the integrals in eq.~\ref{eqn:betti} - \ref{eqn:betach} were defined according to the physical interpretation of connected regions and holes as being neutral regions or ionized bubbles for ionization field. In this paper we aim to maximize the extraction of information contained in the 21cm brightness temperature field and hence the upper and lower limits have been taken in a similar fashion for the $\delta T_b$ field. Therefore, for connected regions $u_{low}=\bar{f}$ and for holes $u_{high}=\bar{f}$, where the overbar represents the mean value of the field $f$ . As we shall see in subsequent sections, such an integration still encapsulates the expected physical interpretation as EoR progresses while employing the complete information contained in the $\delta~T_b$ field.

In addition to the above measures of morphology, in this work we also use area of connected regions and holes described by $A^{ch}_{\mathrm{x}}$, defined analogous to $r^{ch}_{\mathrm x}$ in eq~\ref{eqn:rch} above. Note that unlike \textit{r}, which is defined using eq.~\ref{circle} as the radius of the largest circle having the same perimeter as that of the curve, here we construct $A^{ch}_{\mathrm{x}}$ using the actual area of the individual structures. Being a two dimensional set, area is expected to give more information than the size information contained in $r^{ch}_{\mathrm{x}}$. Therefore , we use a combination of  $A^{ch}_{\mathrm{x}}$ and $r^{ch}_{\mathrm x}$ in conjunction with $\beta^{ch}_{\mathrm x}$ and $N_{\mathrm x}$ for both connected regions and holes of the 21cm brightness temperature field to maximize our information content. 
\section{21cm simulation}
\label{sec:3.1}
The 21cm is the wavelength corresponding to the energy difference between the hyperfine levels of neutral hydrogen in ground state. This energy difference corresponds to a transition temperature of $T_*= 0.068$~K or a frequency of $1420~\mathrm{MHz}$. 
The redshifted frequency lies in the radio range of frequencies so that in the Rayleigh Jean's regime the intensity of this spectral line can be quantified by the brightness temperature $T_b$. This is observed as an offset from the CMB temperature and is called the differential brightness temperature $\delta T_{b}$. For an observed frequency $\nu$, corresponding to a redshift z and at a given point in space $\mathbf{x}$ \cite{Furlanetto:2006jb}:

\begin{equation}
\delta T_b(\nu,\mathbf{x}) \approx 27 \ x_{HI}(\mathbf{x})\left(1+\delta_{nl}(\mathbf{x})\right)\bigg( 1-\frac{T_\gamma(z)}{T_S(\mathbf{x})}\bigg) \ \frac{\Omega_bh^2}{0.023}
\bigg({\frac{1+z}{10}\ \frac{0.15}{\Omega_Mh^{2}}}\bigg)^{1/2} (\rm{mK}), 
\label{eqn:Tb}
\end{equation}
where $\delta_{nl}$ is the evolved density contrast and $T_{\gamma}$ is the CMB temperature. The effect of peculiar velocities is ignored in the above expression because they are very small in magnitude compared to the expansion rate of the universe at the redshifts and scales of interest \cite{Raghu_sim}. Here, $T_S$ is the spin temperature which describes the relative population of the hyperfine levels and is given by the Boltzmann distribution, $n_{1}/n_{0}=3~$exp~$(-T_{*}/T_{S})$. For our study, we work in the post heating regime where $T_S \gg T_{\gamma}$. This assumption is valid during most of the EoR \cite{miralda} as the IGM would have heated by X-ray sources prior to ionization \cite{xray,lya} and $T_S$ would have completely coupled to the gas kinetic temperature of the IGM. Therefore from eq. (\ref{eqn:Tb}) the morphology of $\delta T_b$ would depend solely upon the morphology of $\delta_{nl}$ and $x_{HI}$. 

In order to generate 21cm maps from the EoR we used the publicly available semi-numerical code \texttt{21cmFAST v1.3} \cite{Mesinger:2010ne}. The code generates Gaussian random initial density field and then evolves it using first order perturbation theory (Zel'Dovich approximation \cite{Zeldovich:1969sb}). In order to generate the $x_{HI}$ field, the code uses the excursion set approach pioneered by \cite{Furlanetto:2004nh}. In this approach a central pixel is marked as ionized if the evolved density field when smoothed at progressively smaller scales, starting at a maximum scale of $R_{mfp}$ meets the following condition for the ionizing efficiency $\zeta$ at some scale $R$:
\beq
f_{coll}(\mathbf{x},z,R) \ge \zeta^{-1}.
\label{eq:3.2}
\eeq
Here $f_{coll}$ is the collapsed fraction for a minimum mass $M_{min}$ required for a halo to collapse.
 The minimum mass, $M_{min}$ is usually taken to be the virial mass $M_{vir}$ corresponding to which the temperature $T_{vir}$ is the temperature above which atomic or molecular hydrogen cooling can occur within a halo.

\section{Constructing Mock 21cm data} 
\label{mock}
Our mock 21 cm data includes model 21cm maps from the EoR and system noise from the observing telescope. For our current analysis we have ignored the effects of foreground contamination and would explore it in a future work. In this section we describe the various steps involved in constructing the mock observational data.


A radio interferometeric observation measures a quantity called the visibility, $\mathcal{V}(\vec{U},\nu)$ which is essentially a 2 dimensional inverse Fourier transform of the sky brightness in a particular direction, $I_{\nu} (\vec{\theta})$ convolved with the beam response function $ A(\theta)$:
\beq
\mathcal{V}(\vec{U},\nu)=\int d^2 ~\theta ~I_{\nu}(\vec{\theta}) ~ A(\theta)~e^{i2\pi \theta.\vec{U}}.
\label{eq:4.1}
\eeq
The visibility depends upon the baseline vector, $\vec{U}=\vec{b}/\lambda$ where $\lambda$ is the observed frequency and $\vec{b}$ is the vector between two antenna elements.  The plane of the antenna baseline vectors is called the \textit{uv}-plane, where \textit{u} is the east-west direction and \textit{v} is the north-south direction. The \textit{uv}-plane is sampled according to the baseline distribution of the antenna array. Therefore, some information of the signal is lost while obtaining the inverse Fourier transform. The actual signal visibility, $S(\vec{U},\nu)$ is therefore the product of the baseline sampling function in the \textit{uv}-plane and the true signal visibility. This is called the \textit{dirty image}. The actual measured visibility $\mathcal{V}(\vec{U},\nu)$, would also have a noise component $N(\vec{U},\nu)$ in addition to the \textit{dirty image} signal component $S(U,\nu)$ in eq.~\ref{eq:4.1}. The final observed image would then be the inverse Fourier transform of $ S(\vec{U},\nu)+N(\vec{U},\nu)$. Unlike statistical studies in the Fourier space, while performing real space analysis it is essential to smooth our field to avoid effects of shot noise. Usually the smoothing scale would be greater than the spread in the convolving function $A(\theta)$ in eq.~\ref{eq:4.1}. Therefore, the inverse Fourier transform of $V(\vec{U},\nu)$ is the smoothed $I_{\nu}(\vec{\theta})$, which is related to the brightness temperature $\delta T_b$ as:

\beq
I_{\nu} (\vec{\theta})=\cfrac{2k_B \nu^2}{c^2}~\delta T_b(\vec{\theta},\nu).
\label{eq:4.2}
\eeq
Therefore, the two important components for modelling mock observations are the baseline distribution of the interferometer and noise sensitivity of each baseline. 
In this study, we consider the noise configuration from SKA I-low as it will have the required sensitivity to image the Epoch of Reionization, as mentioned earlier.  In the following subsections we describe the signal and noise map used for constructing mock $\delta T_b$ maps for our analysis.
The box size used for this analysis is larger than in our earlier studies. This is because a bigger box size mitigates uncertainities due to cosmic variance inherent in the analysis. 
\vskip -6.cm 
\subsection{Signal Map}
\label{sec.:6.1}
In order to construct our mock EoR signal map, we generated the $\delta T_b$ field at a redshift of $z=7.4$  on a $512^3$ grid of a $400 ~ \rm Mpc$ box using $21 \mathrm{cmFASTv1.3}$. This gives a pixel resolution of $\sim 0.78 ~ \rm Mpc$. The initial conditions were generated on a $1024^3$ grid at redshift $z=300$.
We choose to work with $\zeta=17.5$, $T_{vir}=3 \times 10^4$ K and $R_{mfp}$ is set to $20$ \rmfamily{Mpc}.The model results in the end of reionization at $z_e\sim6$ and the mean neutral hydrogen fraction, $\mathrm{x}_{HI}=0.5$ at $z=7.4$. Hereafter, the input model will be referred to as the \textit{fiducial} model.  We choose $z=7.4$ because it gives a good compromise between the noise rms of SKA and the level of physical information that can be extracted from our statistics.

The size and resolution of the box along the line of sight fixes the bandwidth and frequency resolution of our observations respectively. The same two quantities in the tangential direction, quantify the maximum field of view and angular resolution.   The signal maps in two-dimensions are constructed by averaging $\delta T_b$ along the line of sight over a  thickness corresponding to a frequency channel width of $\Delta \nu_c = 1~\mathrm{MHz}$.  The $400$ Mpc box used in this study corresponds to a maximum angular scale of $\theta_{max}\sim 2.55^{o}$ or a minimum baseline of $U_{min}\simeq 22.7$ at the observed frequency of $169 \mathrm{MHz}$. The resolution of the simulation grid is $0.78$ \textrm{Mpc} and corresponds to an angular resolution of $\Delta \theta \simeq 0.3'$ or a maximum baseline of $U_{max} \simeq 11455$.

\begin{figure}[H]
	\centering
	\includegraphics[height=5.8cm,width=7.2cm]{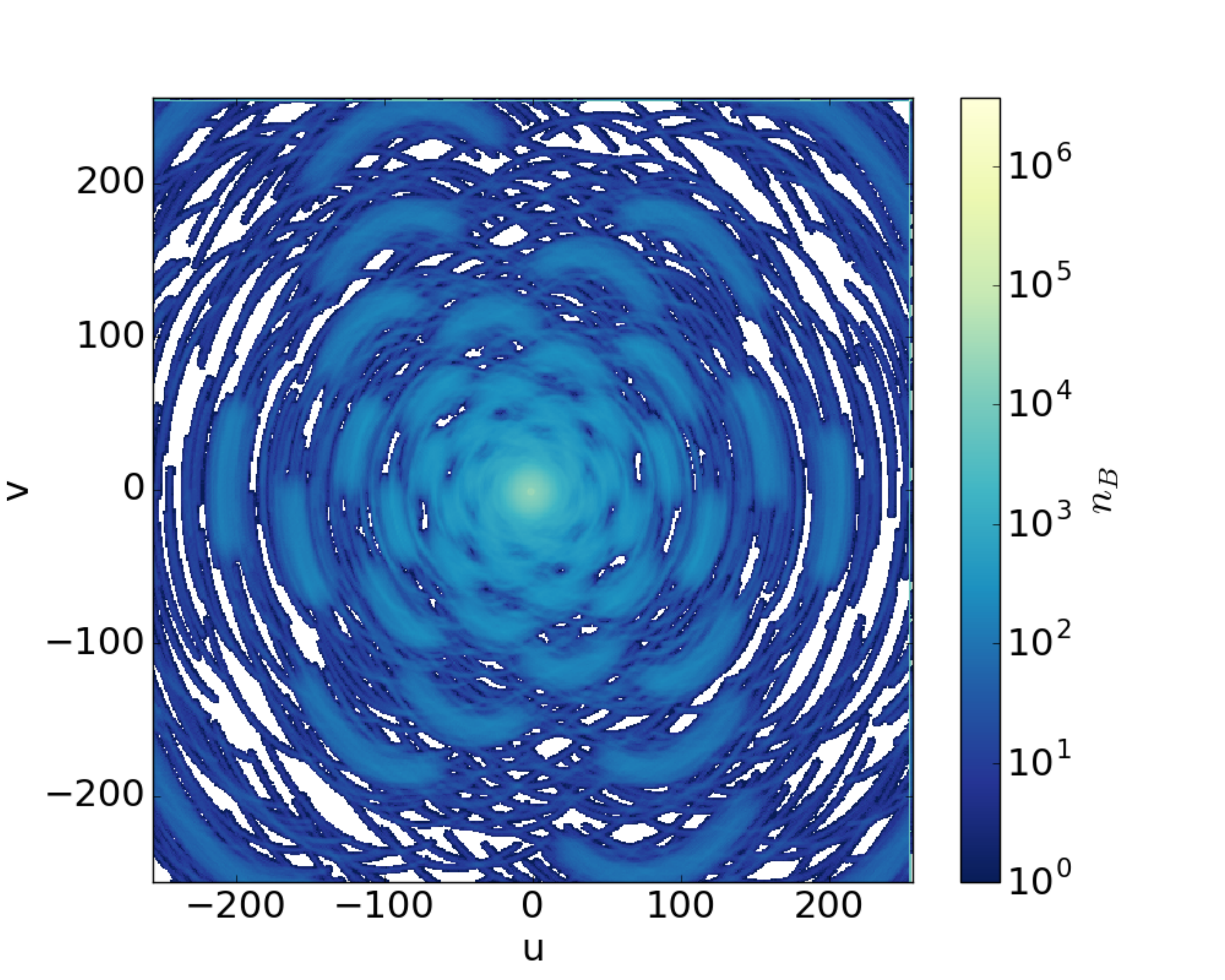}
	\caption{The \textit{uv}-coverage of the SKA-I low for 4 hours of observation per day at a declination of $\delta_{dec}=-30^{o}$ , Right Ascension of $0^o$ and integration time $\Delta t_c=120$ seconds.}
	\label{uv}
\end{figure}
In order to obtain the \textit{dirty image}, we generate the two-dimensional baseline distribution, $n_B^{i,j}$ in a 512$\times$ 512 grid for a 4 hours per day observation with an integration time of $\Delta t_c=120$ seconds. The contribution of galactic nonthermal emission would add to the system noise hence the observations must be carried out at higher galactic latitudes. Therefore the southern galactic pole at around a declination of $-30$ is considered for our mock observations. In fig.\ref{uv} we plot the baseline distribution used for our mock observation.
\begin{table}[H]
	
	\centering 
	\begin{tabular}{c c}
		
		\hline
		\hline                        
		SKA Parameter & Value \\[0.5ex]
		\hline
		\hline				Redshift ($z$)& 7.4 \\
		Central Frequency ($\nu_c$)& 169 $\mathrm{MHz}$\\
		Frequency Resolution ($\Delta \nu_c$) &1 $\mathrm{MHz}$\\
		Integration time ($\Delta t_c$)& $120~\mathrm{sec}$\\
		System Temperature ($T_{sys}$) & $100+60\times(300~\mathrm{MHz/\nu_c})^{2.55}$~K \\
		Number of Antenna($N_{ant}$) & 512\\
		Effective Collecting area ($A_{eff}$) & $962~\mathrm{m^2}$\\ 
		
		\hline
		\hline
		
	\end{tabular}
	\caption{The SKA I instrument parameters used in this study for generating our mock observation at $z=7.4$.}
	\label{Table:1}		
	
\end{table} 

In table \ref{Table:1} we list the SKA parameters used for this study \footnote{\url{https://www.skatelescope.org/wp-content/uploads/2012/06/84\_rsm-v1.0-word-1.pdf}}. After generating the \textit{uv} coverage, the following steps have been adopted for constructing the \textit{dirty image}:
\begin{itemize}
	\item Generate the $\delta T_b$ maps corresponding to a bandwidth of $1~\mathrm{MHz}$ from 21cmFAST using our \textit{fiducial model} and take a two dimensional Fourier transform.
	\item Incorporate the effect of empty pixels in the \textit{uv} coverage by including a mask, such the the value at $i,j$th pixel is 0 where there are no baselines and 1 where $n_B^{i,j} \ne 0$.
	\item Multiply the mask by the Fourier transformed map and take an inverse fourier transform of the product to obtain the \textit{dirty image}.
\end{itemize} 
We find that the \textit{dirty image} map is visually hardly distinguishable from the signal map as the \textit{uv} coverage is almost filled. We had also found that at high smoothing scales (that have been employed for this study), the morphology of dirty image and signal map are almost equal.


\subsection{Noise maps}
The noise from two different baselines is uncorrelated. We model $N(\vec{U},\nu)$ in the \textit{uv}-space and take it's inverse Fourier transform to obtain the real space noise. This real space noise is then added to the \textit{dirty image} of the signal map, in order to obtain our mock image.

The RMS brightness temperature sensitivity for each baseline for a frequency channel width $\Delta \nu_c$ and correlator integration time $t_c$ for an observed frequency $\nu$ is given by \cite{Thompson}:
\beq
\sigma_N=\cfrac{c^2~T_{sys}}{\nu ^2 A_{eff}~\Delta \Omega ~\sqrt{2~\Delta \nu_c \Delta t_c}}~,
\label{eq:4.3}
\eeq
where $A_{eff}$ is the effective collecting area of each antenna, $\Delta \Omega= (\Delta \theta)^2$ is the beam solid angle, $T_{sys}$ is the system temperature and $c$ is the speed of light. We have chosen $\Delta t_c=120$ seconds. We carry out the following steps to generate the noise maps \cite{ghara2017}:
\begin{itemize}
	\item The $512 \times 512$ Fourier space grid (or \textit{uv}-space) is populated with a Gaussian random field with mean zero and standard deviation $\sigma_N$ given in eq. \ref{eq:4.3}.
	\item Since a point on the \textit{uv}-grid can correspond to multiple baselines, the noise in the (\textit{i,j}) th pixel reduces by a factor of $1/\sqrt{n_B^{i,j}}$.
	\item When averaged over long observation hours, the noise will further reduce by a factor of $\sqrt{t_{obs}/t_{obs}^{\textit{uv}}}$, where $t_{obs}^{\textit{uv}}$ is the observation time per day which is $4$ hours for our analyses.
	\item To account for empty pixels in the \textit{uv}- grid a mask is included which is 0 at the empty pixels and 1 otherwise as done before for generating the \textit{dirty image}.
	\item The final real space map is obtained by taking the inverse Fourier transform of this reduced noise.
\end{itemize}
We define the signal to noise ratio in images as:
\beq
SNR_{image}=\cfrac{\mu_{T_S}}{\sigma_{T_N}} ,
\eeq
where $\sigma_{T_N}$ is the standard deviation in the smoothed noise map obtained after performing the steps above and $\mu_{T_S}$ is the mean in the smoothed signal map without masking and noise addition (see section \ref{sec.:6.1}).

\begin{figure}
	\begin{subfigure}{.22\textwidth}
		\centering
		\includegraphics[height=3.7cm,width=4.8cm]{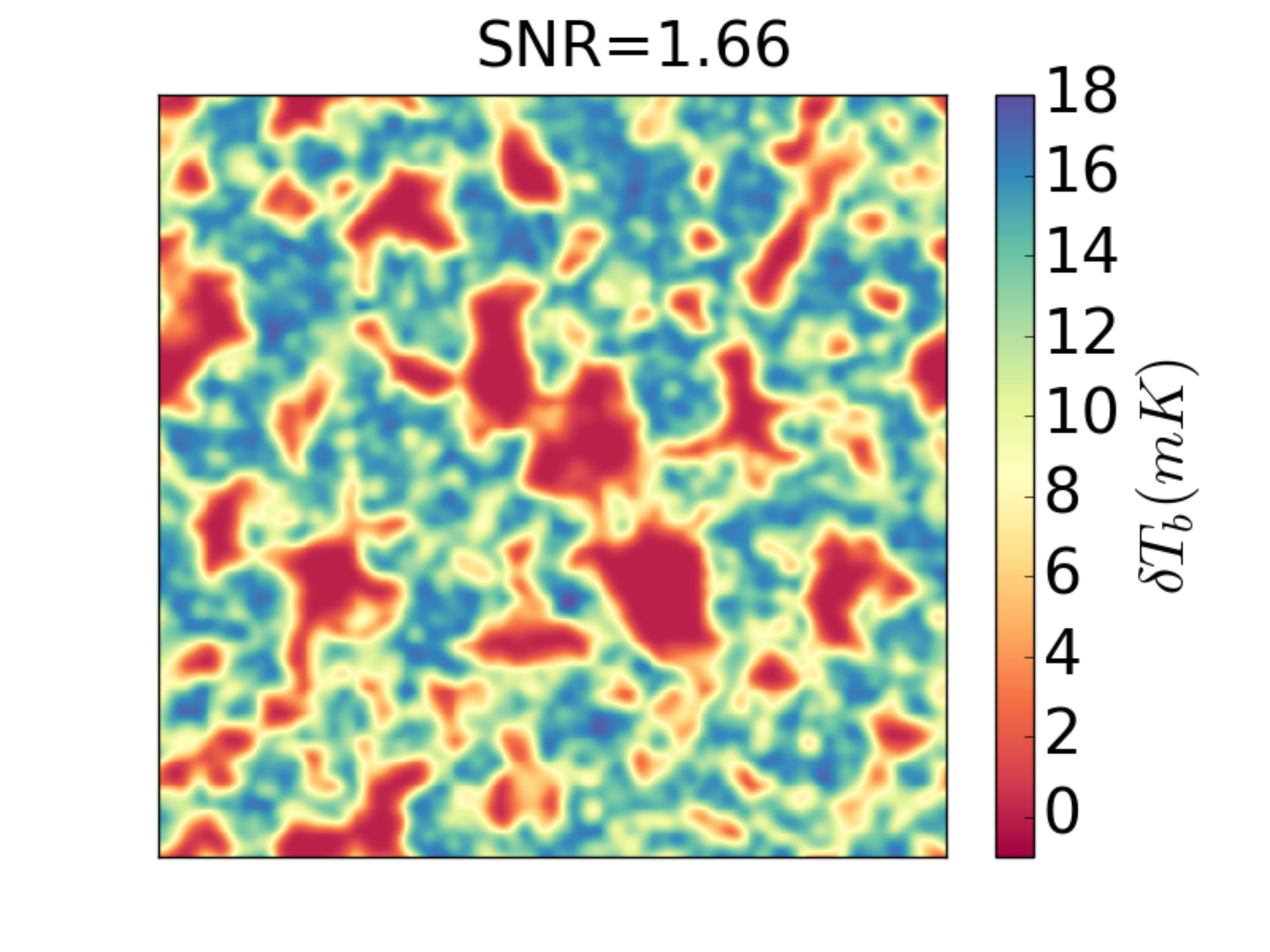}
		\subcaption{}
	\end{subfigure}
	\begin{subfigure}{.22\textwidth}
		\centering
		\includegraphics[height=3.7cm,width=4.8cm]{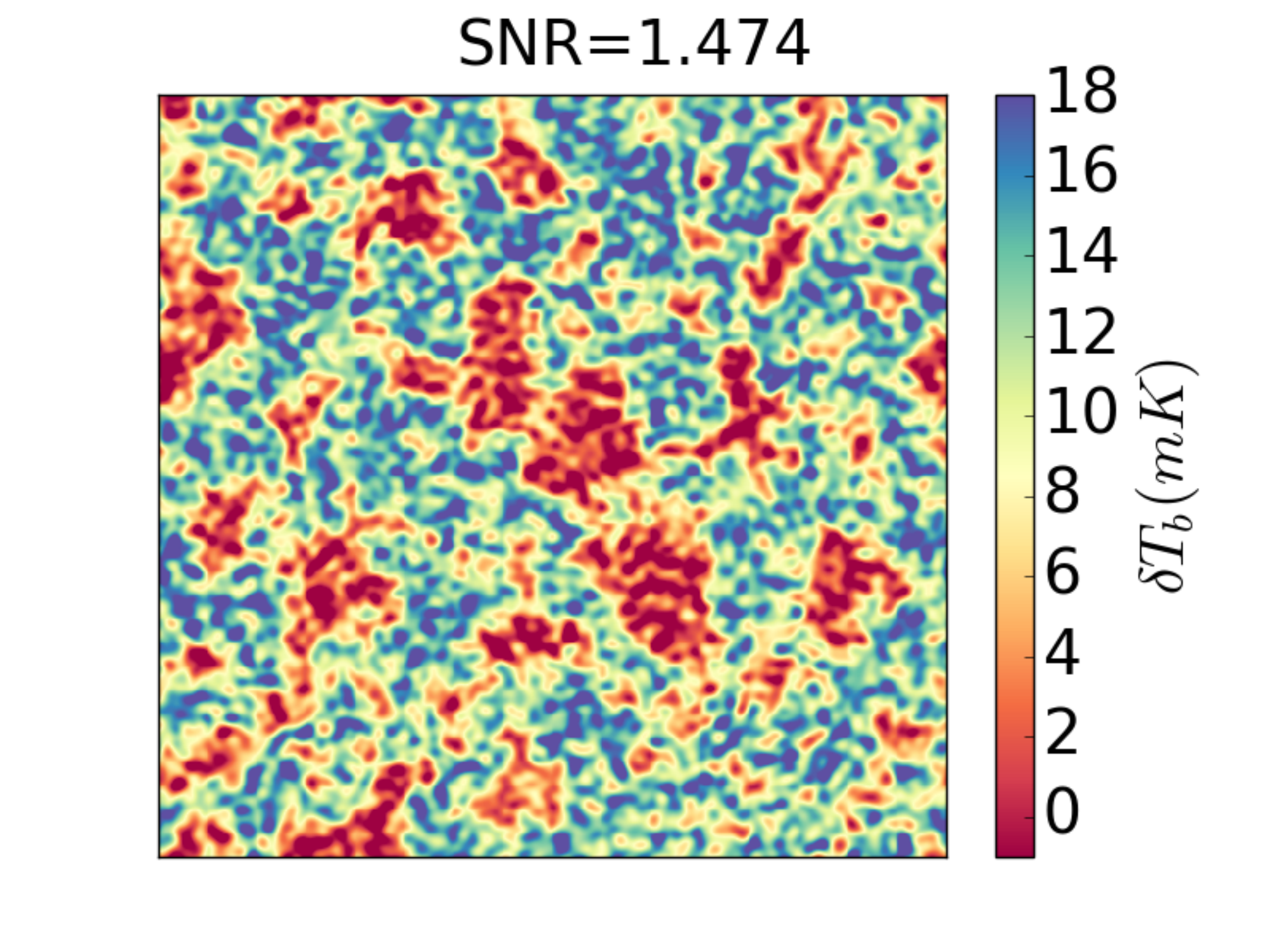}
		\subcaption{}
	\end{subfigure}
	\begin{subfigure}{.22\textwidth}
		\centering
		\includegraphics[height=3.7cm,width=4.8cm]{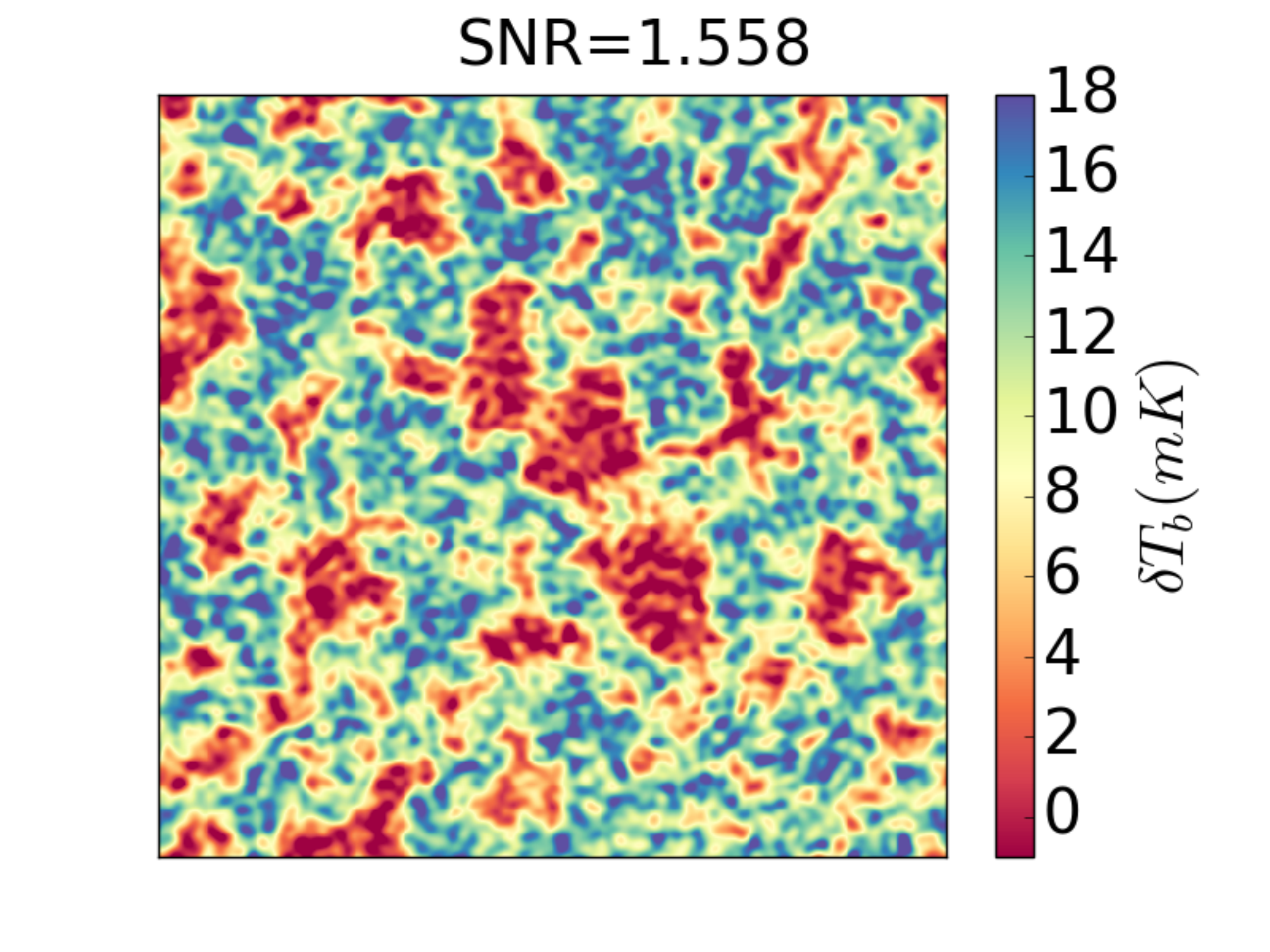}
		\subcaption{} 
	\end{subfigure}
	\begin{subfigure}{.22\textwidth}
		\centering
		\includegraphics[height=3.7cm,width=4.8cm]{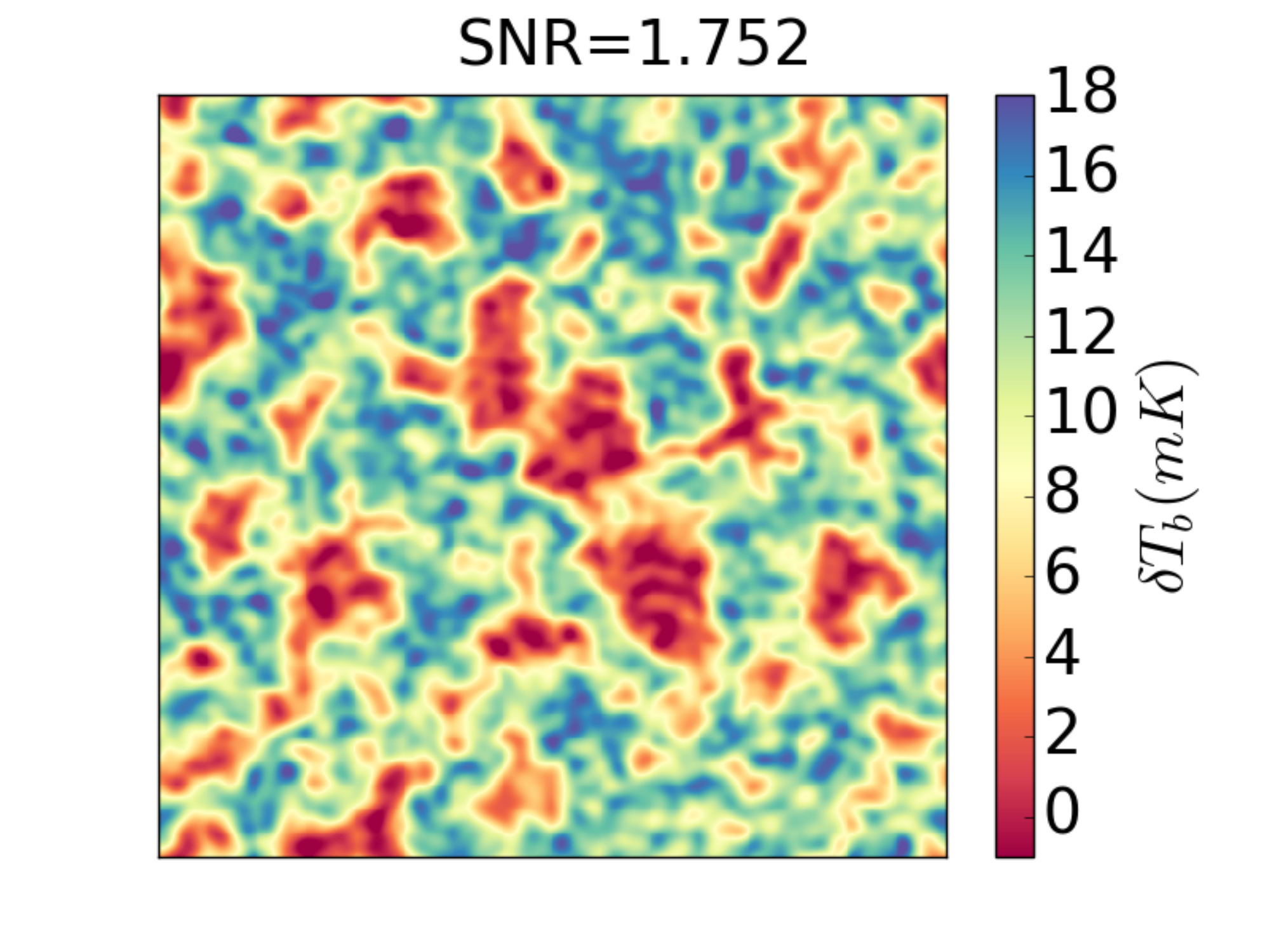}
		\subcaption{} 
	\end{subfigure}
	\caption{Maps showing the effect of noise on image of $\delta T_b$ for the field of view of $2.55^o \times 2.55^o$ corresponding to our $400$ \rmfamily{Mpc} box  and frequency resolution of $1$ MHz. Fig.(a) shows the signal map without masking and noise, smoothed at $R_s=4.5$ Mpc with an SNR=1.66. Fig.(b) shows the map with SKA noise for $t_{obs}=1000$ hrs, $R_s=4.5$ Mpc and SNR=1.474. Fig.(c) shows the noise image for $t_{obs}=2000$ hrs, $R_s=4.5$ Mpc and SNR=1.558. The map in fig. (d) is for $t_{obs}=1000$ hrs, $ R_s=6.5$ Mpc and SNR=1.752.}
	\label{maps}	
\end{figure}

The noise in images can be reduced in two ways. The first way is to increase the time of observation $t_{obs}$ or increase the frequency channel width $\Delta \nu_c$. The second way is to smooth the images or reduce the resolution. In fig.~\ref{maps} we show four $\delta T_b$ maps to compare the effect of noise from SKA I low for the parameters in table \ref{Table:1}. The first map on the left panel shows the signal map without masking and noise addition, smoothed at $R_s=4.5$ Mpc. The second and third map are the mock noisy maps at $R_s=4.5$ Mpc smoothing for $t_{obs}=1000$ and $t_{obs}=2000$ hours, respectively. We find that the SNR in images increases by almost $\sim 5.7 \%$ when  $t_{obs}$ is doubled. The rightmost panel shows the same map smoothed at $6.5$ Mpc for $t_{obs}=1000$ hours. We find that the SNR increases by almost $18.7 \%$  for the same $t_{obs}$ for a $1.5$ times increase in the smoothing scale. The effect of noise is to introduce small scale structures in the image. When smoothed, these small scale structures get washed out and therefore increases the SNR of the image. However, even though smoothing increases the SNR in images much faster than an increase in $t_{obs}$, it also smooths out small scale structures that are intrinsic part of the true underlying signal which leads to a loss of information.

\section{Parameter Space}
\label{parameter}
We work with theoretical models described by varying three parameters -- $\zeta$, $T_{vir}$ and $R_{mfp}$. In this section we will analyze the variation of our statistics in parameter space at a fixed redshift of $z=7.4$. In our earlier studies we focussed on the variation of our statistics with redshift. The understanding gleaned from this study will enable us to anticipate the behaviour of the posterior when performing Bayesian analysis (section \ref{Bayes}) for our mock observation constructed at the same redshift (section \ref{mock}).
We generate ideal simulations in a grid of $15 \times 15 \times 15 $  models spanned by $R_{mfp}, \zeta $ and $T_{vir}$ and calculate our statistics numerically for these models using the method described in section~\ref{sec:2}. The range of the parameters used are: $10~\mathrm{Mpc} \le R_{mfp} \le50 ~\mathrm{Mpc}$, $10 \le \zeta \le 24$ and $1 \times 10^4 ~\mathrm{K} \le T_{vir} \le 8\times 10^4 ~\mathrm{K}$

We use the same box parameters as for our mock signal simulation described in section \ref{mock}. The same set of initial conditions has been used for all models. In order to be consistent with our mock observed maps, the statistics have been calculated as an average over  24 slices, each having a thickness of $16.5$ \rmfamily{MPc} (corresponding to the $1$ \rmfamily{MHz} frequency bandwidth used for constructing the noise map described in section~\ref{mock}). All slices have been smoothed at a smoothing scale of $R_s=9.5 \mathrm{Mpc}$. The reason for the choice of smoothing scale will be described in section~\ref{bias}.
\begin{figure}[H]
	\includegraphics[height=4.5cm,width=4.86cm]{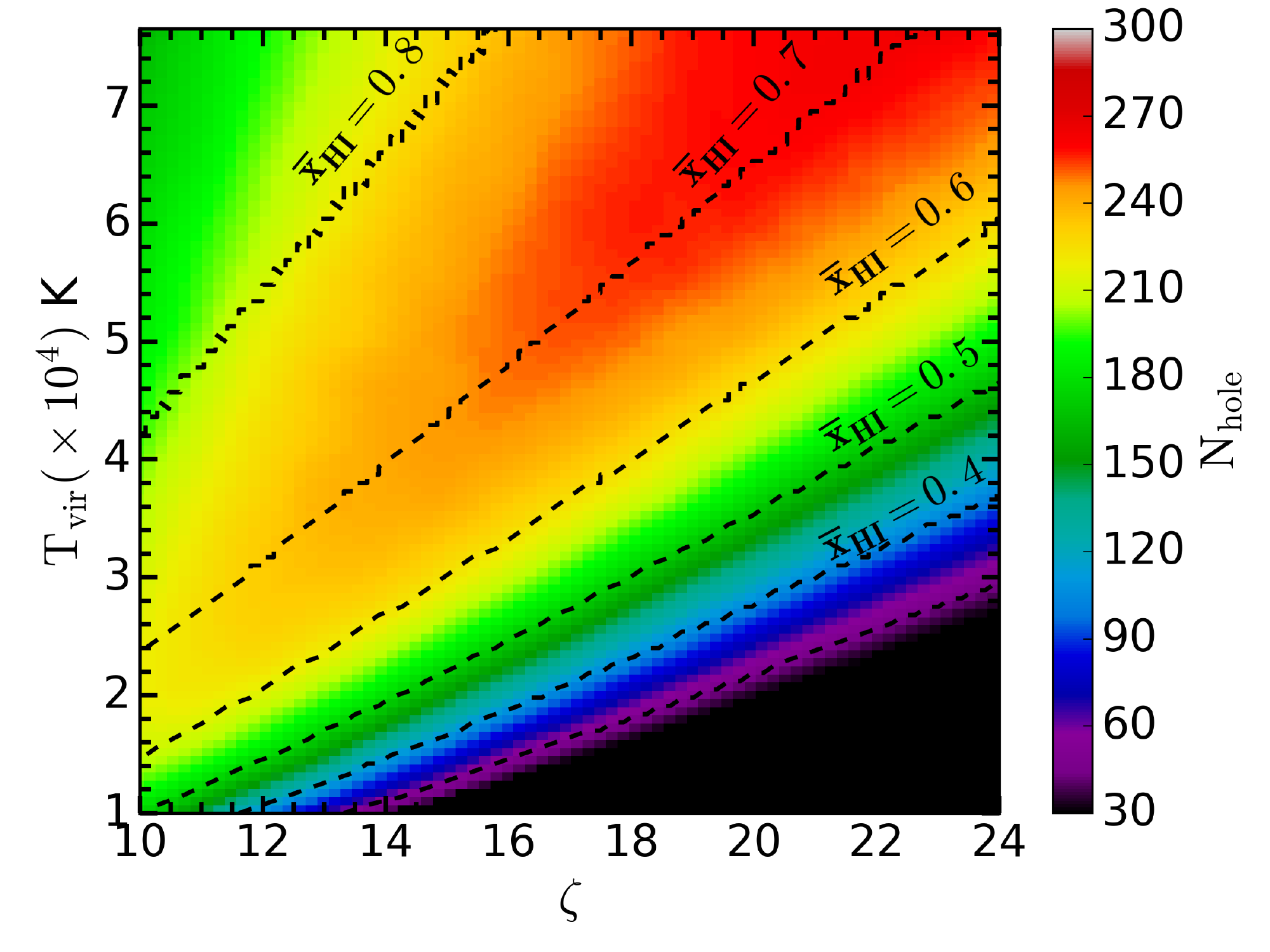}
	\includegraphics[height=4.5cm,width=4.86cm]{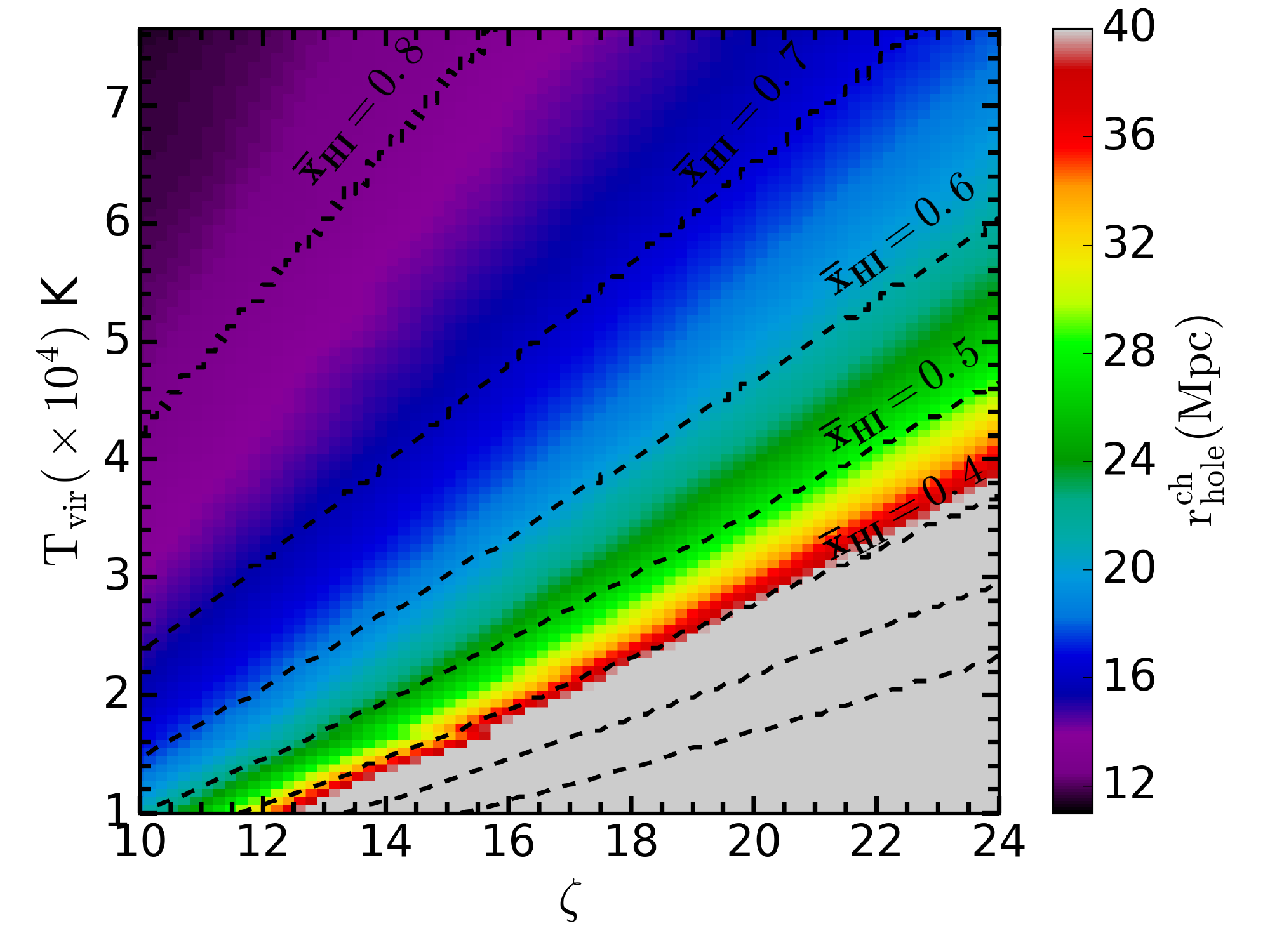}
	\includegraphics[height=4.5cm,width=4.86cm]{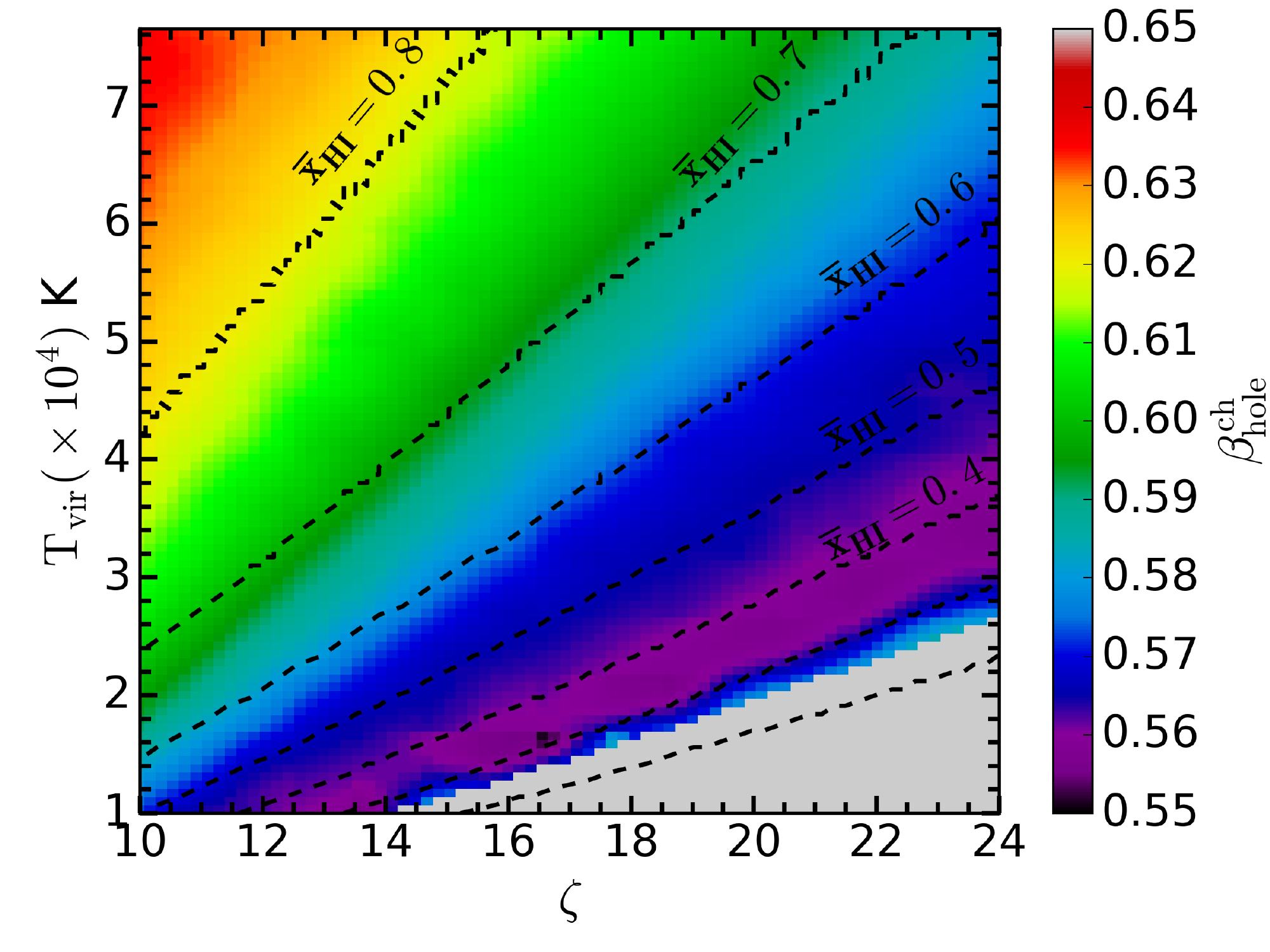} \\
 \includegraphics[height=4.5cm,width=4.86cm]{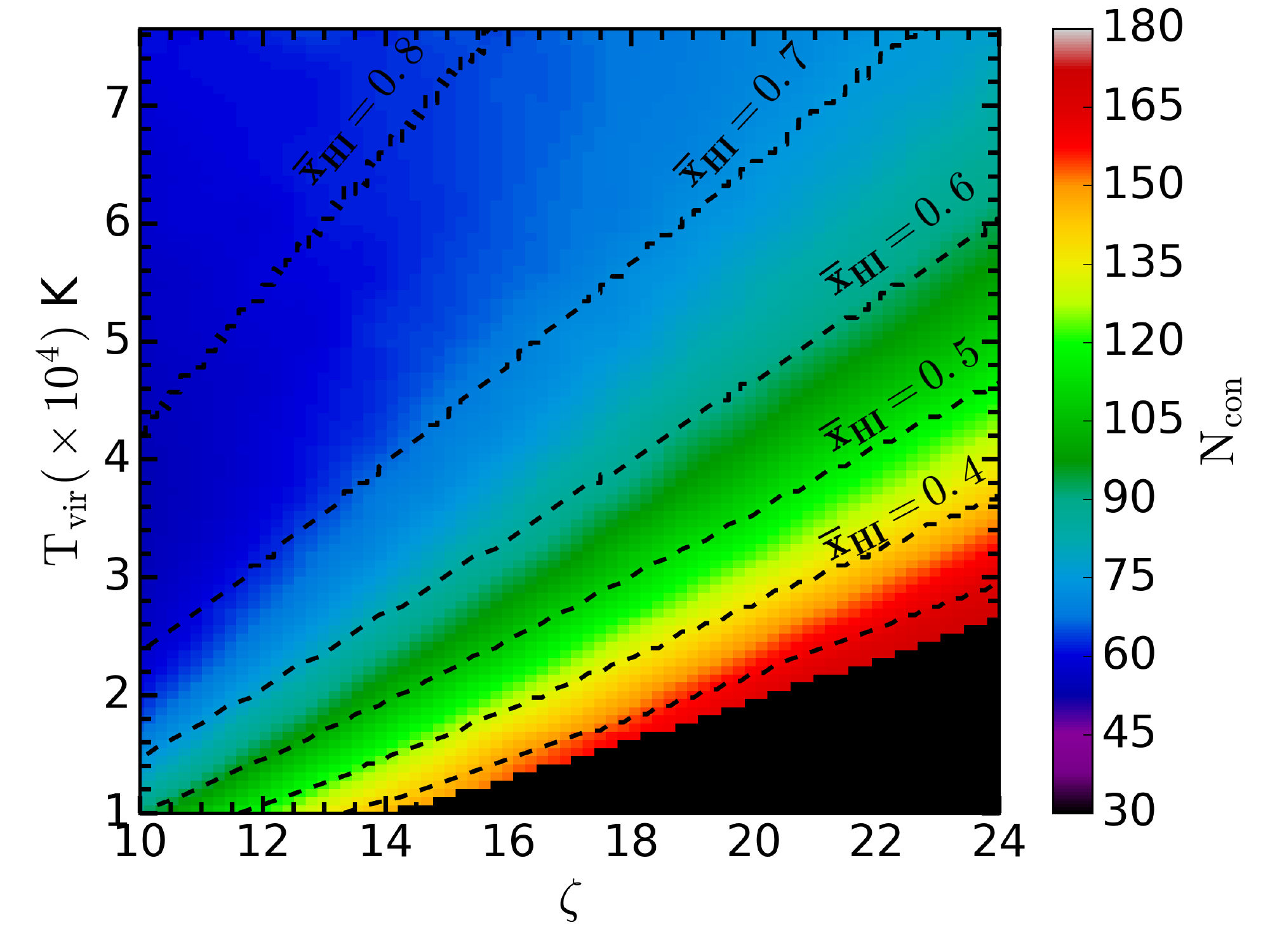}
\includegraphics[height=4.5cm,width=4.86cm]{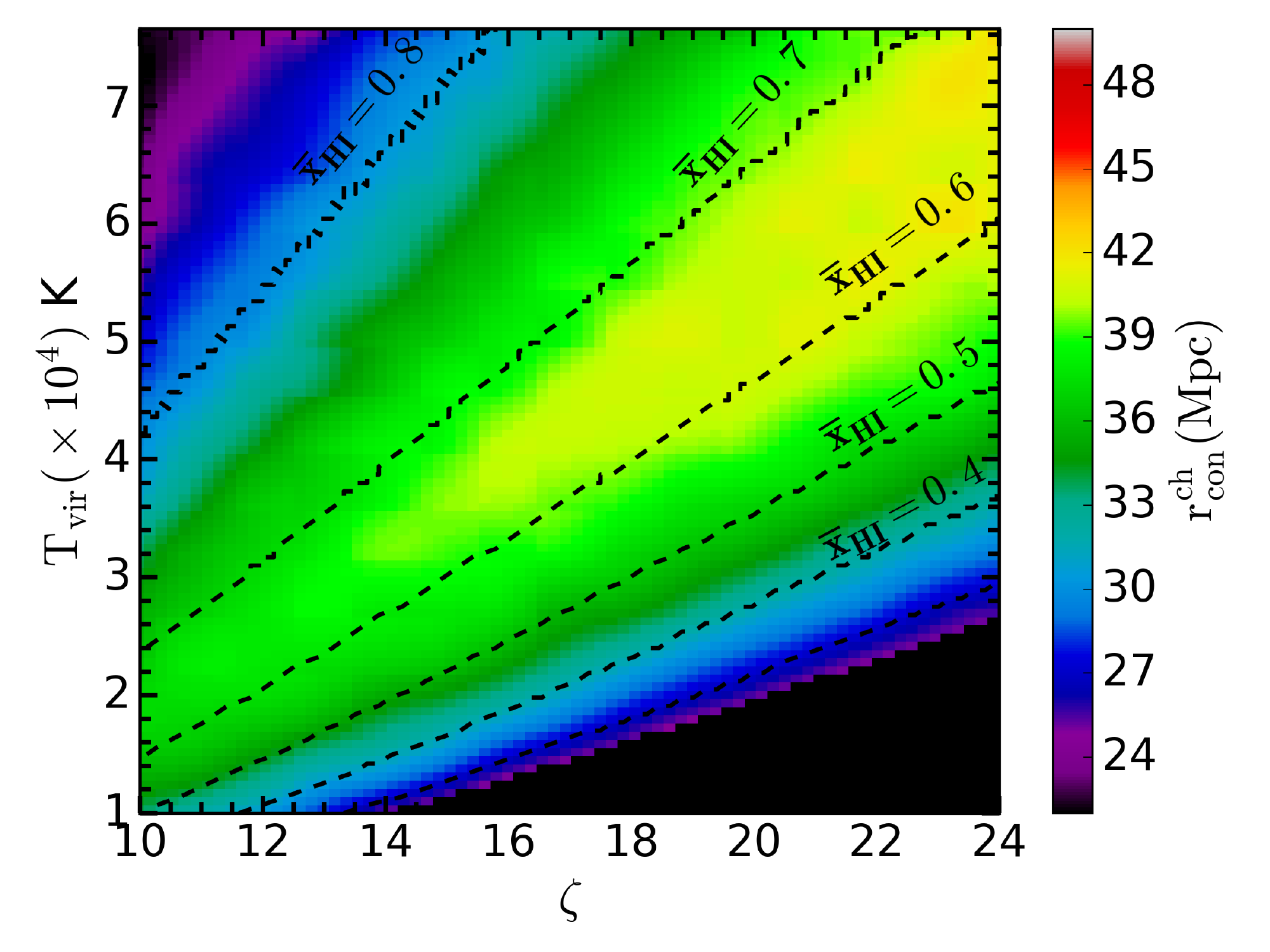}
\includegraphics[height=4.5cm,width=4.86cm]{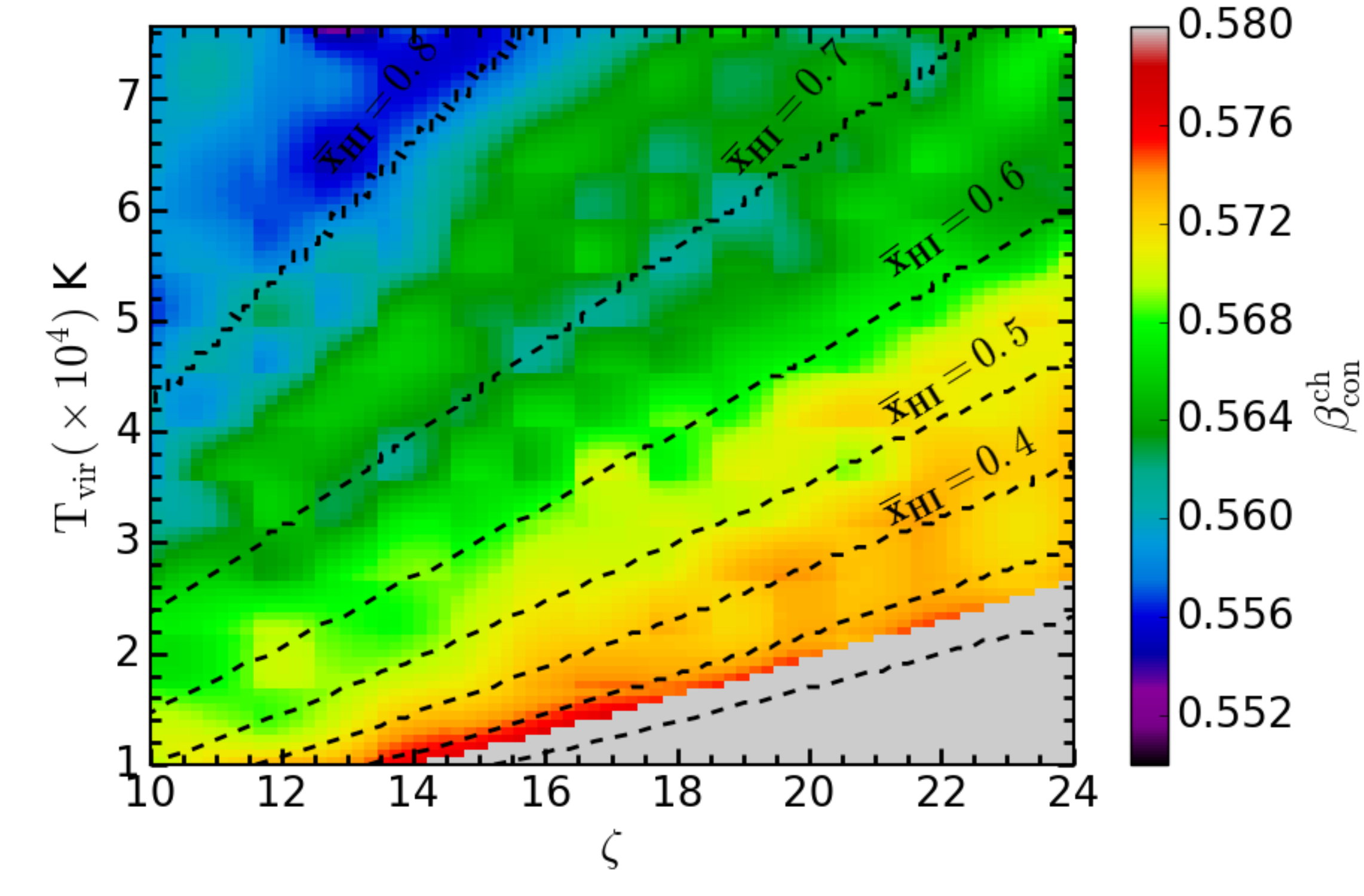} \\

	\caption{The figure shows the behaviour of our statistics in the 2D parameter space spanned by $\zeta-T_{vir}$ for $R_{mfp}=20$ \textrm{Mpc}. The figures are plotted by interpolating $10^4$ points in parameter space over $15 \times 15$ models for which the statistics are calculated numerically using the method described in section~\ref{sec:2}. The colorbar shows the values of the respective statistics and the black dotted lines are lines of constant $\bar{x}_{HI}$. The panels show $N_{\mathrm{x}}$ (\textit{left column}), $r^{ch}_{\mathrm{x}}$ (\textit{middle column})and $\beta^{ch}_{\mathrm{x}}$ (\textit{right column}) where ${\mathrm{x}}=hole$ \textit{(top row)} or $con$ \textit{(bottom row)}.}
	\label{stat2d}
\end{figure}
The color maps in fig.~\ref{stat2d} encapsulate the dependence of the three statistics in the $\zeta-T_{vir}$ plane for $R_{mfp}=20$ \textrm{Mpc}. These color maps show the values of $N_\mathrm{x}$,  $r^{\rm ch}_\mathrm{x}$ and $\beta^{\rm ch}_\mathrm{x}$ obtained by interpolating the values of the respective statistics over the $15 \times 15$ models with values computed numerically using the method described in section \ref{sec:2}. The interpolation errors for sampling in this grid are less than $1\%$ of the exact value within this range of parameters. The top row shows the hole statistics and the bottom row shows the connected region statistics. The dotted lines show the variation of $\bar{x}_{HI}$ in the parameter space. The statistics are smoothly varying functions of the parameters. By visual inspection of the plots, we find that interpolation introduces features only at very high ($>0.8$) and very low ($<0.3$) values of $\bar{x}_{HI}$. Since, each model is at different stages of it's ionization history at $z=7.4$, we can interpret the color maps as a variation with $\bar{x}_{HI}$. The variation with $\bar{x}_{HI}$ is the same as studying the redshift evolution for a fixed model as described in \cite{Kapahtia2019}. We would like to emphasize that the definition of a connected region (or hole) as being a neutral (or ionized region) depends upon the mean value of the field (section \ref{sec:2}). Since a decreasing neutral fraction leads to creation of new ionized regions, we find that while this physical interpretation of holes as being ionized regions is straightforward, this definition of connected regions as being neutral regions is not straightforward to interpret. 
The top left corner of the color maps corresponds to a high value of average neutral fraction, i.e. regions where the morphology is dominated by a big neutral (or connected regions) region dotted with numerous ionized bubbles (or holes).  These bubbles increase in size and number as the big neutral region fragments. Therefore, from around $x_{HI}=0.8$ to $x_{HI}=0.7$, we see an increase in $N_{hole}$ and a corresponding increase in $r^{ch}_{hole}$. Thereafter, there is a sharp increase in size of holes and a corresponding decrease in number of holes, due to mergers of ionized regions. This is reflected in both $N_{hole}$ and $r^{ch}_{hole}$, in going diagonally towards the bottom right region of the parameter space from the line corresponding to $\bar{x}_{HI} \simeq 0.7$. The growing and merging bubbles decrease the value of $\beta^{ch}_{hole}$ and the decrease is faster for values of $x_{HI} \le 0.7$. At very low values of $x_{HI}$, $\beta^{ch}_{hole}$ again abruptly begins to increase indicating completion of ionization. The important point to notice from the figures, is that the values of $N_{hole}$ and $r^{ch}_{con}$ are degenerate for some set of models. We also find that $\beta^{ch}_{\mathrm{x}}$ vary very mildly over the parameter space.

\begin{figure}[H]
	\centering
	\includegraphics[height=10.0cm,width=6.8cm]{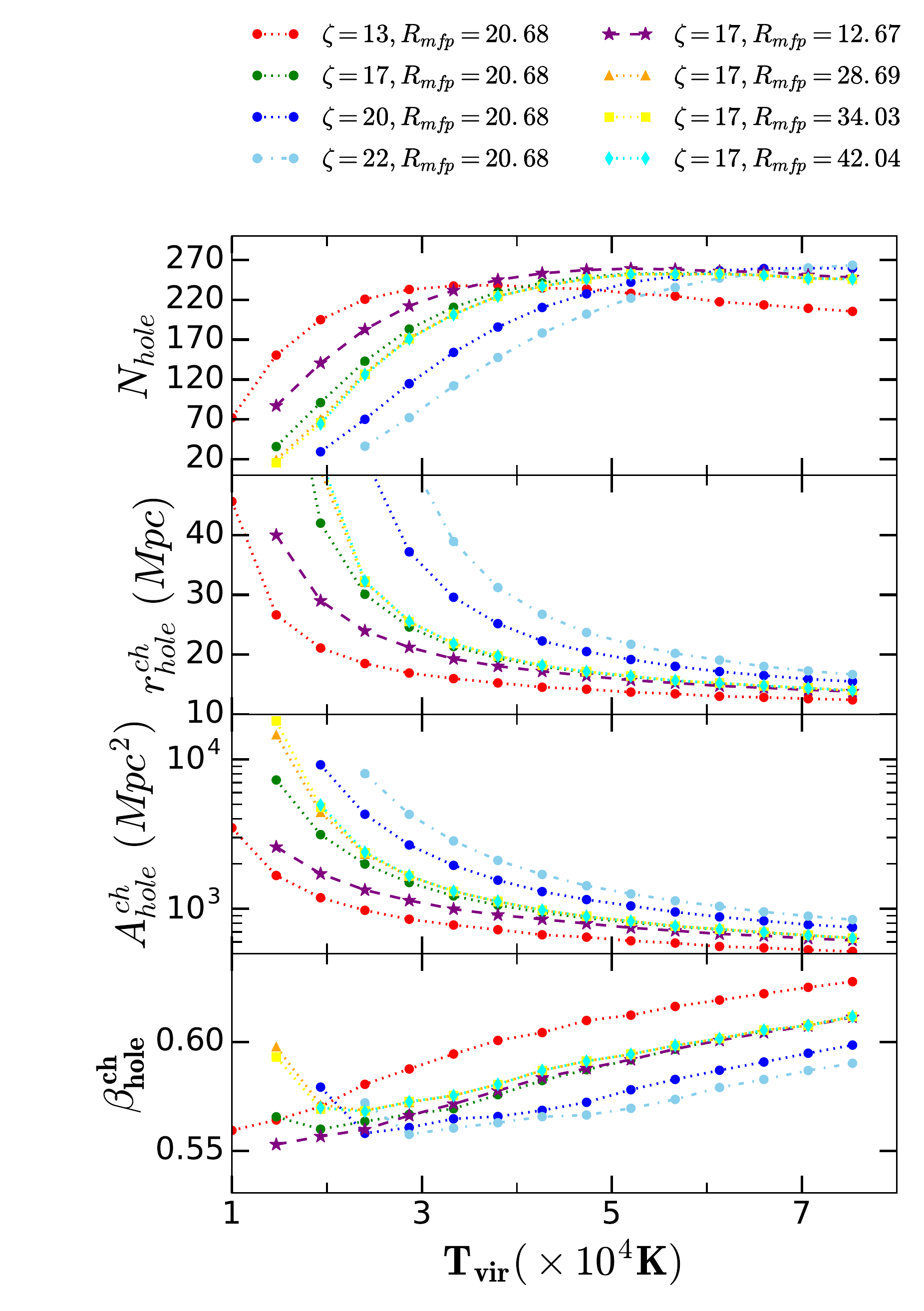} \hspace{1.2cm}
	\includegraphics[height=10.0cm,width=6.8cm]{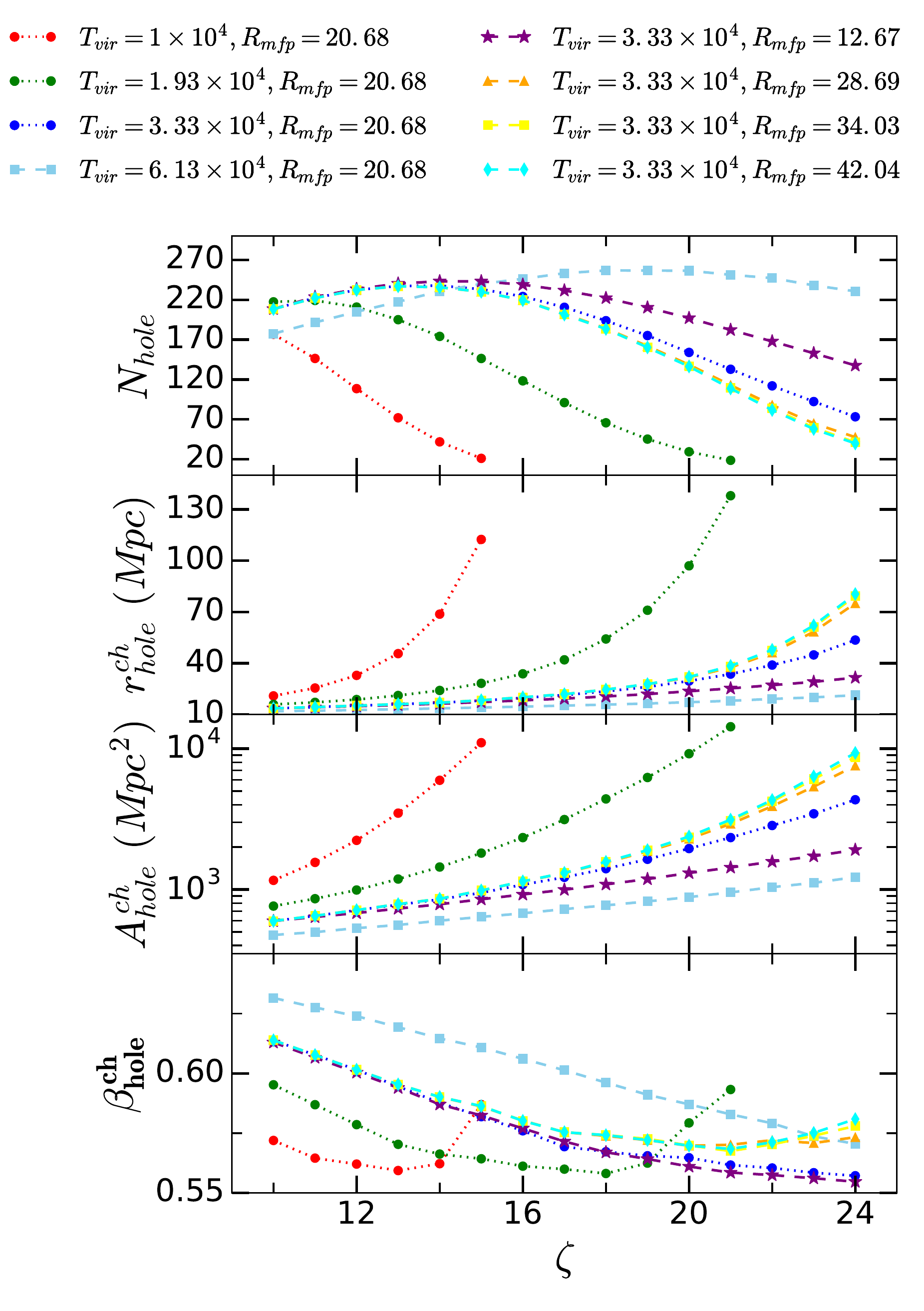}\\ \vspace{1.2cm}
	\includegraphics[height=10.0cm,width=6.8cm]{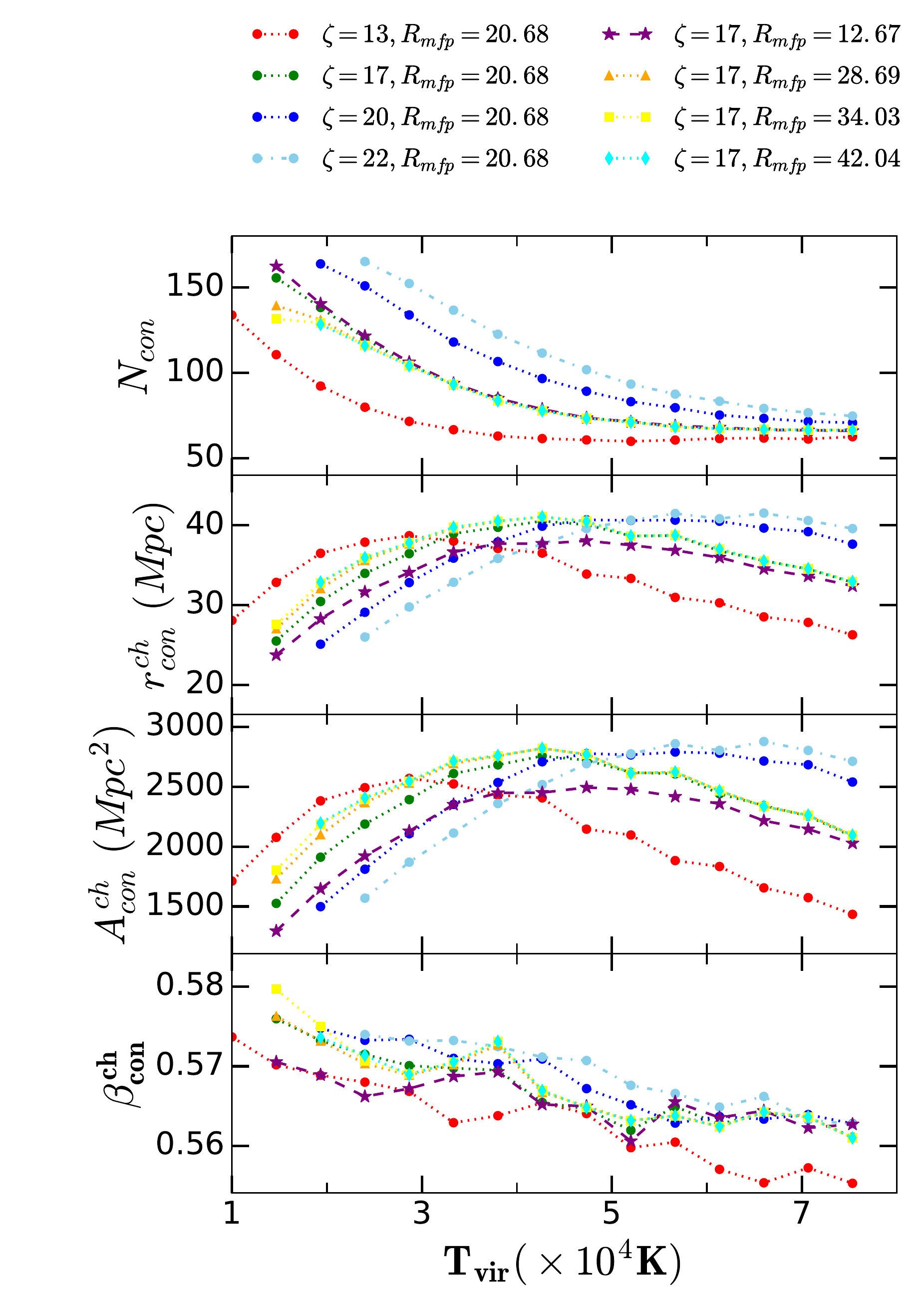} \hspace{1.2cm}
	\includegraphics[height=10.0cm,width=6.8cm]{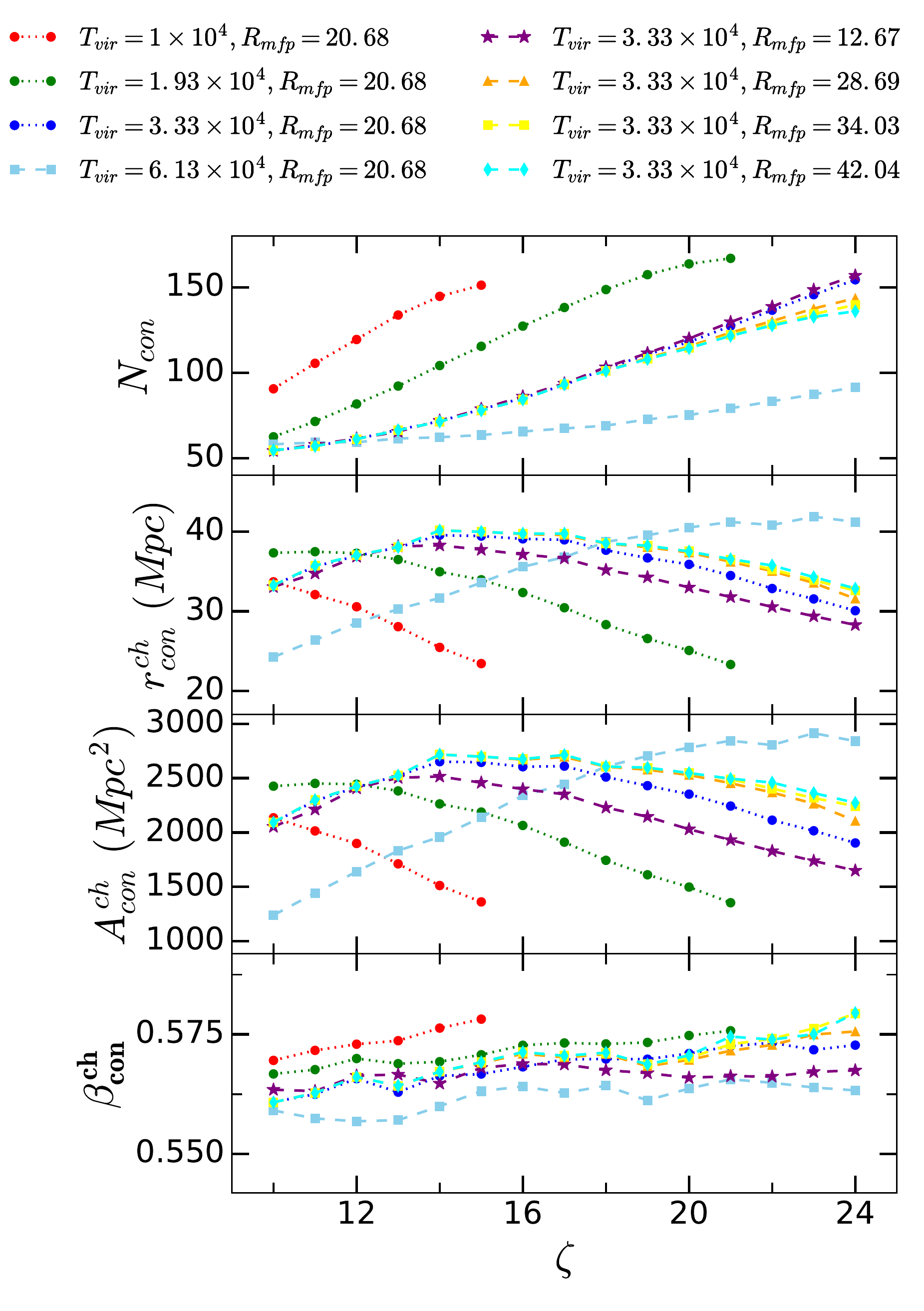}\\
	\caption{The figure shows the variation of hole (\textit{top row}) and connected region (\textit{bottom row}) statistics with $T_{vir}$ (\textit{left column}) and with $\zeta$ (\textit{right column}) for different fixed values of the remaining two parameters. In each plot $N_{\mathrm{x}}$ (\textit{top panel}), $r_{\mathrm{x}}^{ch}$ (\textit{second panel}), $A_{\mathrm{x}}^{ch}$ (\textit{third panel}) and $\beta^{ch}_\mathrm{x}$ (\textit{fourth panel}).}
	\label{stat_fix}
\end{figure}

In fig.~\ref{stat_fix} we show the variation of our statistics with $T_{vir}$ at different fixed values of $\zeta$ and $R_{mfp}$ on the left panel and vice versa on the right panel. The top row shows the hole statistics and the bottom row shows connected region statistics. If the value of $R_{mfp}$ is fixed then for a given $\zeta$, decreasing $T_{vir}$ corresponds to an increasing ionized fraction and same holds for a fixed $T_{vir}$ and increasing $\zeta$. Moreover, if the value of $\zeta$ is higher then complete ionization is obtained at a relatively larger value of $T_{vir}$, since the effect of increasing $\zeta$ is to increase the rate of progress of ionization. Therefore ionization is achieved at a lower collapse fraction (or higher $T_{vir}$) for a higher $\zeta$. This is captured in the left panels of the figure. On the other hand if the collapse fraction is high (or low $T_{vir}$) then ionization is acheived at a lower $\zeta$. This is captured in the right panels. This variation is captured by the red, green, dark blue and skyblue dotted lines. Thus, at high $T_{vir}$ values there are few isolated bubbles, which initially increase as $T_{vir}$ decreases. A further decrease in $T_{vir}$ will lead to more numerous bubbles and increased mergers which decreases the number of holes. Therefore the number of holes first increases and then decreases (see fig.~\ref{stat2d}). In fig. \ref{stat_fix}, this turnover is seen at $\zeta=13$  (shown in red) where the ionization history is varying faster. For other $\zeta$ values this turnover happens beyond the range of $T_{vir}$ values plotted in fig.~\ref{stat_fix}. The turnover is a result of the competition between the rate of appearance of new collapsed sources and the rate of bubble mergers \cite{Kapahtia:2017qrg}. On the other hand, for a fixed $\zeta$ (or $T_{vir}$) and vary $R_{mfp}$, we find that the variation with $T_{vir}$ (or $\zeta$) is almost insensitive at high values of $T_{vir}$ (or lower $\zeta$) corresponding to larger values of $x_{HI}$. A smaller value of mean horizon of ionizing radiation leads to reionization becoming slower. This trend is captured by purple, orange, yellow and cyan markers. This parameter is known to be sensitive only when ionized region sizes approaches $\sim R_{mfp}$ \cite{Mesinger:2010ne}.

 As mentioned in section~\ref{sec:2}, using area information, encapsulated in $A^{ch}_\mathrm{x}$ is expected to improve constraints since it is a two dimensional set. We shall also use area as an additional size statistics. The variation of $A^{ch}_\mathrm{x}$ in the parameter space is identical to the variation of $r^{ch}_\mathrm{x}$ and is shown in \ref{stat_fix} (note that log scale has been used for $A^{ch}_{hole}$). 

The description in this section shows that there is a systematic variation of our statistic with parameters and can therefore enable us to constrain models of the EoR. We also find that the variation is sensitive to the variation of $x_{HI}$. Hence we can also obtain constraints on the ionization history. The question then arises as to what combination of statistics would lead to precise and tight constraints. The degeneracy of models for variation of $N_{hole}$ and $r^{ch}_{con}$ necessitates the use of a combination of both scale information and size information to obtain constraints. The anisotropy information would improve the constraints due to extra information contained in the shape anisotropy parameter which is not captured in the former two statistics. Thus, we  shall use a combination of the three morphological descriptors based on number, size and shape and see if we can use them to constrain EoR models.




\section{Covariance matrix and Statistical Biases} 
\label{bias}
In the previous section, we studied the expected behaviour of our statistics in the parameter space under an ideal case scenario. In this section, we analyse the behaviour of our statistics for the mock observed data and compare it with the noiseless case. This will enable us to decide upon an optimum observational strategy. 

\textbf{Error Covariance Matrix:} The error covariance matrix for our statistics is given by:
\begin{equation} 
C_{ij}= \langle(stat_{i}- \langle stat_{i} \rangle)(stat_{j}- \langle stat_{j} \rangle) \rangle,
\end{equation}
where the angle brackets denote average over realizations and $stat_i=N_\mathrm{x}, r^{ch}_\mathrm{x} , \beta^{ch}_\mathrm{x}$ or $A^{ch}_\mathrm{x}$. There will be two separate sources of errors, one due to cosmic variance and the other due to noise. Therefore the total error covariance matrix will be given by:

\beq
C_{ij}^{tot}=C_{ij}^{signal}+C_{ij}^{noise},
\label{cov}
\eeq
where $C_{ij}^{signal}$ is calculated by taking average over noiseless realizations by taking different initial condition seed values for the 21\textrm{cmFAST} simulations and $C_{ij}^{noise}$ is calculated by taking different noise seed values for some fixed noiseless realization to which noise is added. The errorbars for each statistic is calculated by using the diagonal elements of the error covariance matrix. We find that the cosmic variance is much greater than the variance in noise.

We also find that there are two predominant effects that interplay to affect our observed statistics:
\begin{itemize}
	\item \textbf{Smoothing:} As mentioned earlier, the effect of smoothing is to erase small scale structures. This has a two fold effect on our statistics. Firstly, the number of structures $N_\mathrm{x}$ in the excursion set would decrease and there would be loss of information at smaller scales. For a fixed box size this would also translate into increased cosmic variance. Secondly, smoothing would increase the average size of structures, $r^{ch}_\mathrm{x}$ in the excursion set. 
	\item \textbf{Noise:} The effect of noise is to introduce several small scale structures in the excursion set. This increases $N_\mathrm{x}$ in the map and decreases the mean size of structures and hence average $r^{ch}_\mathrm{x}$ is decreased. 
\end{itemize}
\begin{figure}
	\centering
	\includegraphics[height=10.8cm,width=6.5cm]{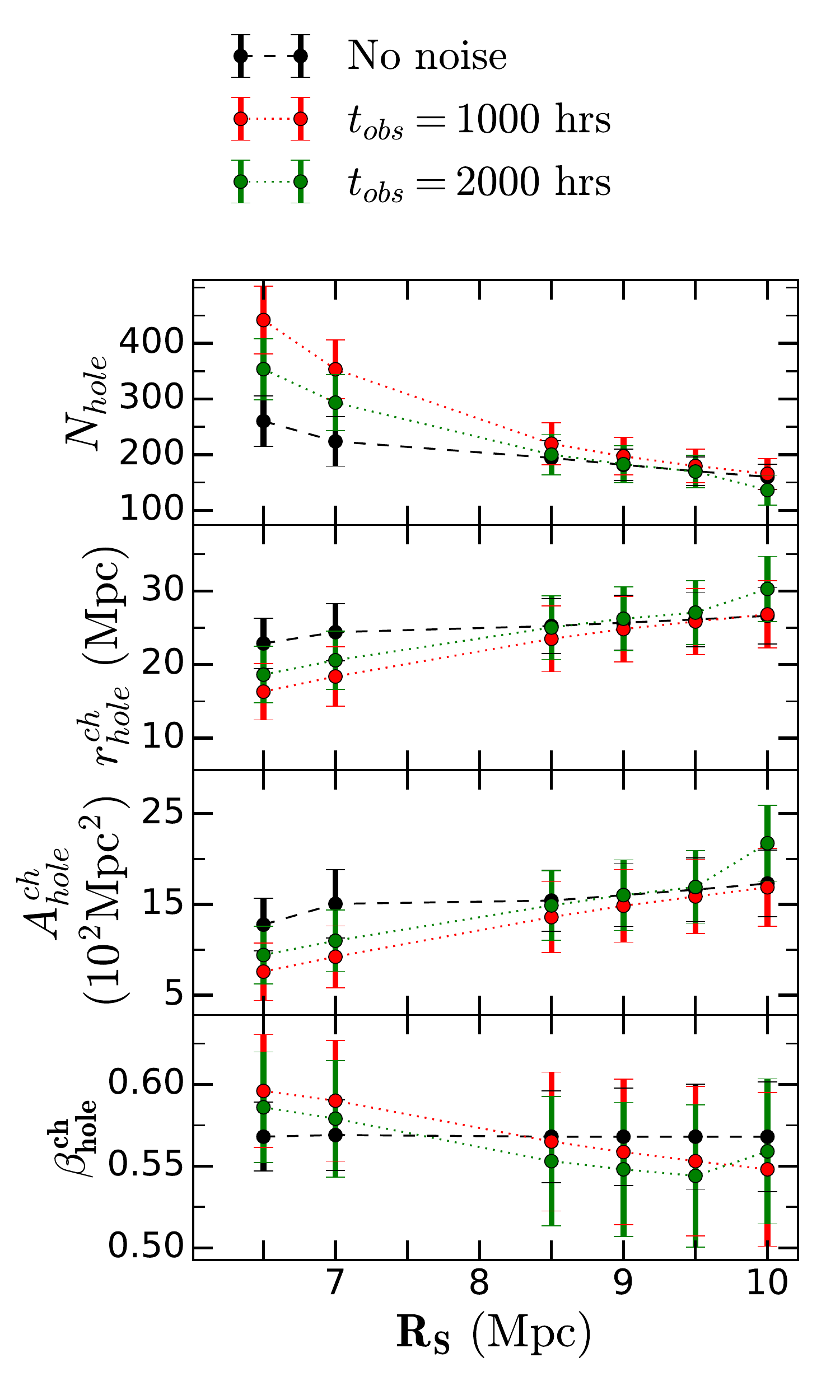} \hspace{2.cm}
	\includegraphics[height=10.8cm,width=6.5cm]{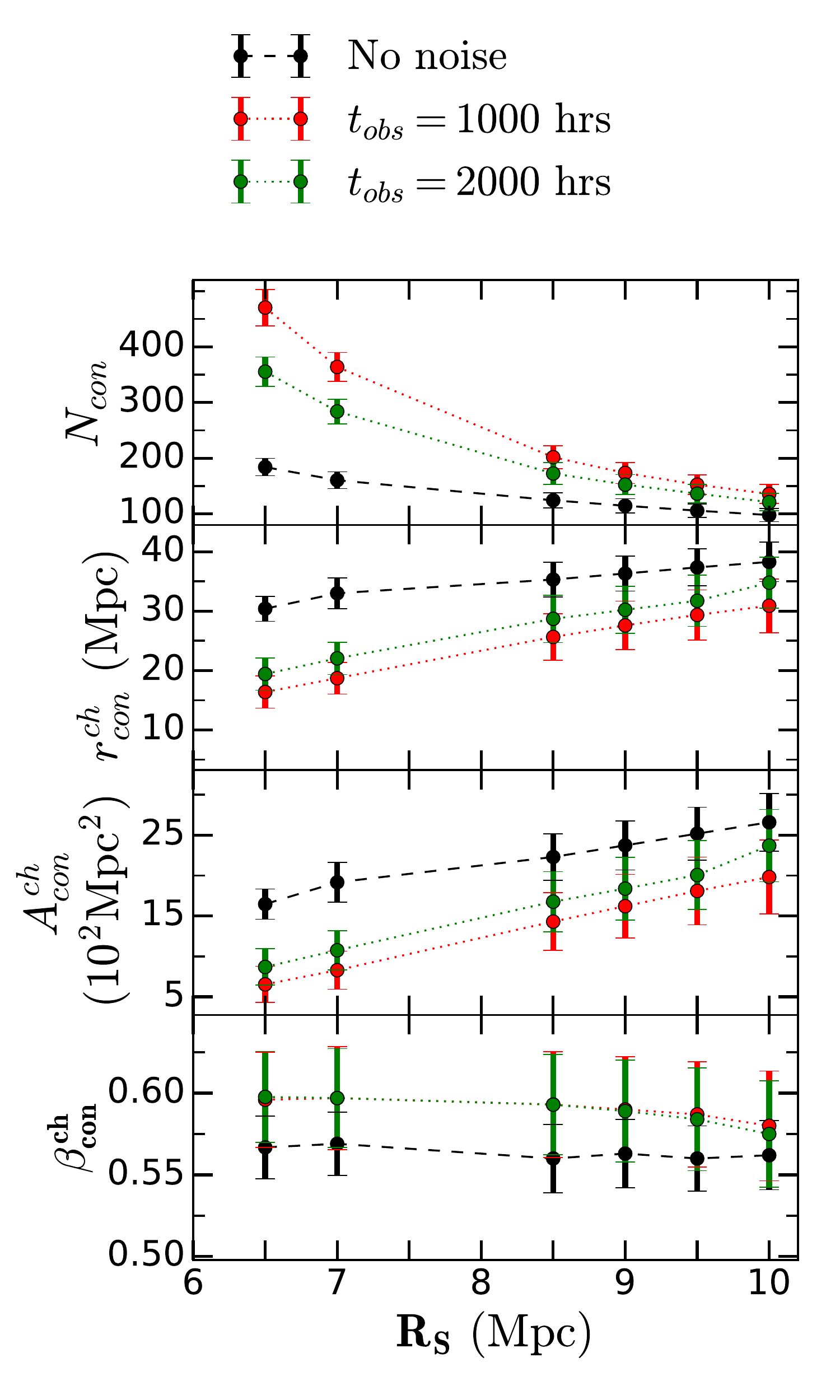}
	\caption{The figure describes the effect of smoothing and noise addition on our statistics. We plot the variation of hole (\textit{left panel}) and connected region (\textit{right panel}) statistic with smoothing scale, $R_s$ for the case of noiseless map (\textit{black dashed line}), noisy map with $t_{obs}=1000$ \textrm{hrs} (\textit{red dotted line}) and $t_{obs}=2000$ \textrm{hrs} (\textit{green dotted line}). The values (and the errorbars) are obtained as a mean over 40 independent realizations.}
	\label{vs_nu}
\end{figure}

Clearly, both smoothing and noise addition would shift the mean value of our statistic. This is shown in fig.~\ref{vs_nu} which shows the variation of our statistic with  smoothing scale for the noiseless case and for an observation time of $1000$ hours and $2000$ hours for our \textit{fiducial} model. We find that in the presence of noise, only when smoothed at a higher smoothing scale, are the mean values closer to the mean for the noiseless case. Also note that there is an increase in the size of errorbars due to addition of noise.

 The errorbars in fig.~\ref{vs_nu} indicate that the noisy case for $t_{obs}=2000$ hours lies within $1-\sigma$ error of the noiseless case at $R_s=9.5~ \mathrm{Mpc}$. Note that the error in noiseless case represents cosmic variance. Therefore, we will carry out our analysis at a smoothing scale of $9.5~\mathrm{Mpc}$. We re-emphasize that a high smoothing scale would also introduce increased statistical uncertainty due to cosmic variance since the number of structures in the excursion set would decrease. Therefore, a bigger box size (or field of view) would be required for our analysis if noise bias is not minimized or corrected and instead choose to reduce noise by smoothing. Also a higher observation time would generate a large amount of data. Therefore, one needs to develop ways to minimize noise in 21cm images, so as to exploit the full signal information embedded in a noisy map. 

\section{Methodology for Bayesian analysis}
\label{Bayes}
In this section, we describe our method of performing a Bayesian analysis to recover constraints on model parameters using our statistics. Since the underlying model for our observed mock simulation is known, the analysis is aimed at showing how well the model is recovered by our statistics. The model used for constructing the mock data is the \textit{fiducial} model described in section \ref{sec.:6.1} with $R_{mfp}=20~\mathrm{Mpc}$, $\zeta=17.5$ and $T_{vir}=3\times 10^4$ K. 

In Bayesian analysis one calculates the posterior probability distribution of model parameters, given the observed value of our statistics. Let $\vec{\theta} = (\zeta,T_{\rm vir}, R_{mfp})$ denote the parameters vector. Let superscript `th' refer to theoretical parameter and `data' refer to (mock) observed data. Let $\vec{X} = stat_i$ denote the vector of the 3 statistics which are functions of $\vec{\theta}$. Then, 
the posterior probability for the model parameters, given the data $\vec{X}^{\rm data}$ is:
\begin{equation}
P(\vec{\theta}^{\,\rm th}|\vec{X}^{\rm data}) = \frac{\mathcal{L}(\vec{X}^{\rm data}|\vec{\theta}^{\rm th})\, \pi(\vec{\theta}^{\rm th})}   {\rm Norm}.
\end{equation}
$\pi$ is the prior probability which we assume to be uniform within the specified prior range.
$\mathcal{L}$ is the likelihood function which we assume to be Gaussian in $\vec{X}$ and for a given covariance matrix $C_{ij}$ it is written as:
\begin{eqnarray}
\mathcal{L}(\vec{X}^{\rm data}|\vec{\theta}^{\rm th}) &=& \frac{1}{\sqrt{(2\pi)^d{\rm det(C)}}}\,\exp\left[-\frac12 \sum_{ij} \Delta X_i\, C^{-1}_{ij}\, \Delta X_j \right] \nonumber\\
&\propto& e^{-\chi^2/2},
\end{eqnarray}
where $\vec{\Delta X} \equiv \vec{X}^{\rm data} - \vec{X}^{\rm th}$, $\chi^2=\sum_{ij}\Delta X_i\, C^{-1}_{ij}\,\Delta X_j$ and $\vec{X}^{\rm th}$ is the theoretically calculated value of the statistic at $\vec{\theta}^{~\rm th}$.  The normalization is given by
\begin{equation}
{\rm Norm} = \int {\rm d}\theta^{\rm th}\, \mathcal{L}(\vec{X}^{\rm data}|\vec{\theta}^{~\rm th})\, \pi(\vec{\theta}^{~\rm th}) ,
\end{equation}
and $\chi^2=\sum_{ij}\Delta X_i\, C^{-1}_{ij}\,\Delta X_j$. 
$\mathcal{L}$ implicitly depends on $T_{\rm vir}$ , $\zeta$ and $R_{mfp}$. The factor $\pi(\vec{\theta}^{\rm th})/{\rm Norm}$ is only a constant which can be calculated to fix the amplitude of the posterior. Therefore,
\begin{equation}
P(\vec{\theta}^{\,\rm th}|\vec{X}^{\rm data}) = \frac{1}{\cal N} \exp\left({-\frac12 \sum_{ij}\Delta X_i\, C^{-1}_{ij}\,\Delta X_j}\right),
\end{equation}
where
\begin{equation}
{\cal N} = {\rm Norm} \times \sqrt{(2\pi)^d{\rm det(C)}}. 
\end{equation}

Thus, in order to obtain the posterior, $P(\vec{\theta}^{\,\rm th}|\vec{X}^{\rm data})$ we need the mock observed value of our statistic, the covariance matrix and the likelihood function.  
We perform our analysis using COSMOMC \cite{cosmomc} as a generic sampler with Gaussian likelihood function. The code performs a  Markov Chain Monte-Carlo sampling of the parameter space to calculate the Likelihood and subsequently the posterior. 
We adopt the following steps for performing the Bayesian analysis:
\begin{enumerate}
	\item We construct 40 independent realizations of our mock data with and without noise, using the parameters and methods described in section \ref{mock}. For noiseless case we construct realizations by varying the initial condition seed of our semi-numerical simulations and for the noisy case we fix a noiseless realization and vary the seed value for the Gaussian random noise added to it, for a $\sigma_N$ calculated using eq.~\ref{eq:4.3}. Thereafter, we use these to calculate the error covariance matrix as described in eq.~\ref{cov}.
	\item We fix our prior range between $10^4 \le T_{vir} \le 8 \times 10^4$ \rmfamily{K}, $10 \le \zeta \le 24$ and $10 \le R_{mfp} \le 50$. 
	\item Due to computational limitations, the Likelihood calculation is performed by interpolating values of the statistic in the parameter space by using the grid of $15 \times 15 \times 15$ models within the prior range, for which the value of the statistics is calculated numerically as described in section \ref{parameter}. As stated before, the errors due to interpolation are $\lesssim 1\%$ and vary across the parameter space. Therefore, the optimal error in interpolation of a given descriptor is chosen as the most probable value of error from the distribution of errors in the parameter space and is added to the covariance matrix.
	\item All maps are smoothed at a scale of $R_S\simeq 9.5$ \textrm{Mpc} in order for the signal to dominate the noise for the observation times under consideration as shown in fig.~\ref{vs_nu} and described in the previous section. 
\end{enumerate}

\section{Results}
\label{results}

\begin{figure}
 	\includegraphics[height=10.cm,width=14.cm]{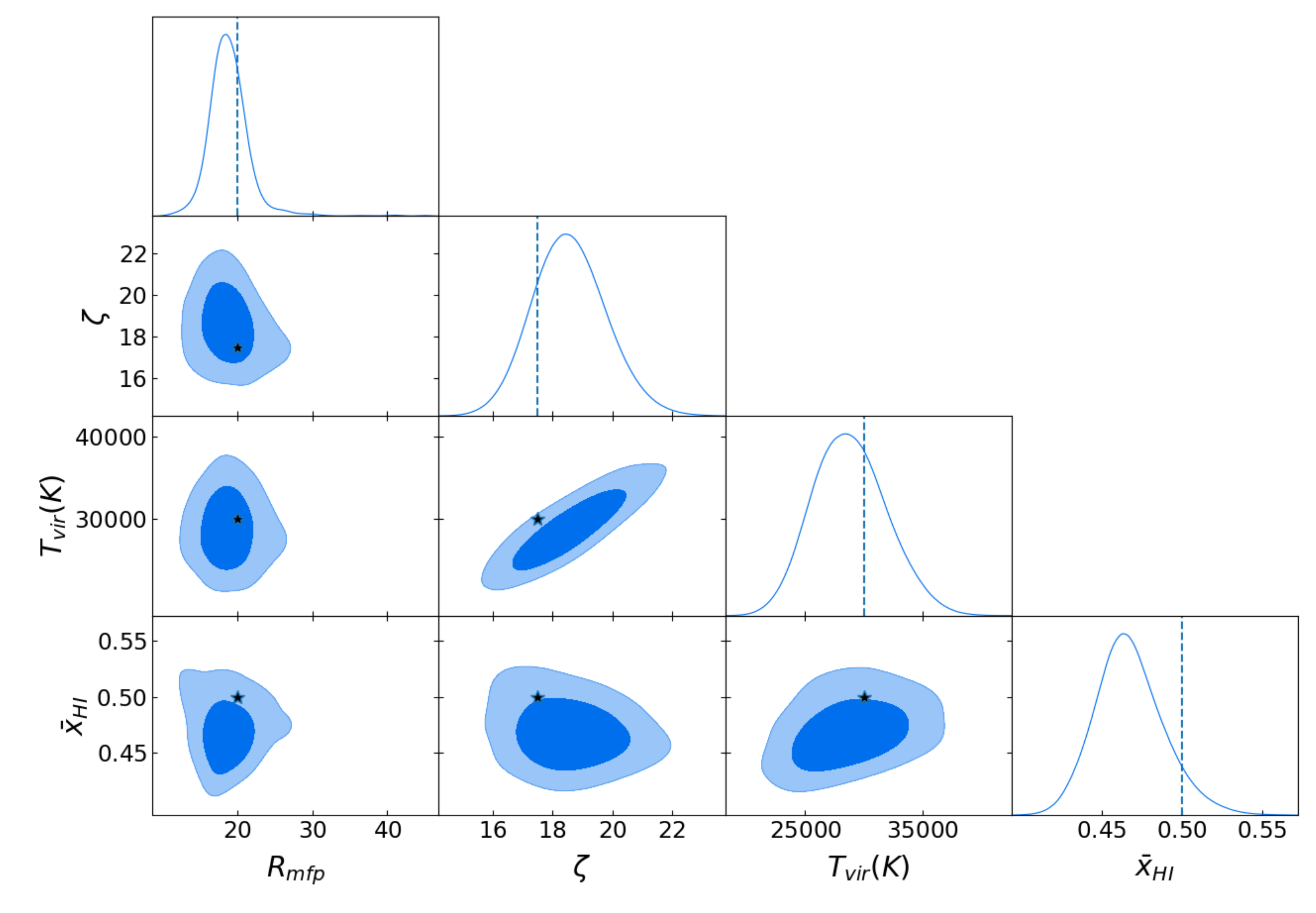} 

 	\caption{The plot shows the posterior probability distribution for a noiseless mock $\delta T_b$ image at a redshift of $z=7.4$, smoothed at $R_s=9.5~\mathrm{Mpc}$.The black star shows the input mock model ($\zeta=17.5, T_{vir}=3\times 10^4~\mathrm{K}$ and $R_{mfp}=20~\mathrm{Mpc}$). } 
 	\label{ell_nl}
 \end{figure}

In this section we present the results of the inference methodology described in previous sections. The triangle plot in fig.~\ref{ell_nl} shows the posterior probabilities for the three parameters $R_{mfp}$, $\zeta$ and $T_{vir}$ for a noiseless simulation map smoothed at $R_s \simeq 9.5$ \textrm{Mpc}.  The input model of our mock observation is marked by the black star and the vertical dashed lines correspond to the input model parameter values. The corresponding table for the best fit values are shown in the top panel of table~\ref{table_nl}. We quantify the tightness of our best fit values by the Coefficient of Variation(CV) which is defined as the ratio of the the standard deviation to the best-fit value. In order to quantify the accuracy of the best fit values we use the relative difference between the best fit and the input model parameters. 
The triangle plots also show the constraints on $\bar{x}_{HI}$, obtained by treating it as a derived parameter.
 We find that we are able to recover our input model to within $68 \%$ credible interval. The values in the bottom panel of table~\ref{table_nl} are the best fit values and the $68\%$ credible intervals of our statistics, corresponding to the best-fit values of the parameters above.
 
 We find that the recovered best fit values have almost similar accuracy for $R_{mfp}$,  $\zeta$ and $T_{vir}$ while the constraints are tighter for $\zeta$ and $T_{vir}$ as compared to $R_{mfp}$. The less tight constraints are expected from the mild variation of $R_{mfp}$ as ionization progresses. The ellipses are shown for a single realization of the observed data. On changing the realization we find similar behaviour. The fourth row of the triangle plot shows the ellipses for $\overline{\mathrm {x}}_{HI}$.  We find strong constraints on ${\mathrm{\bar{x}}_{HI}}$ to within $68 \%$ credible intervals. Even though the recovered best fit values of ${\mathrm{\bar x}_{HI}}$ are not very accurate relative to those on the model parameters, the constraints on best fit are tighter ($CV \simeq 0.05$) than the parameter constraints. The mild variation of the statistics with $R_{mfp}$ (as shown in fig.~\ref{stat_fix}) and the strong correlation between $\zeta$ and $T_{vir}$ which is almost along the lines of constant $\mathrm {\bar x}_{HI}$ (fig.~\ref{maps}), makes these statistics a promising tool to constrain the ionization history of the universe.

 The physical interpretation of the values of the statistics in the bottom panel of table~\ref{table_nl} rely on the mean value of the field which in this case roughly corresponds to $\bar{\mathrm{x}}_{HI} \simeq 0.5$ (also recovered from the best fit value of $\bar{\mathrm{x}}_{HI}$). As shown in fig.~\ref{stat_fix} , we expect strong degeneracy in the value of the statistics for $R_{mfp}$, while the correlation of the statistics shows opposite behaviour with $\zeta$ and $T_{vir}$.  In the fourth column we see the ellipses for the statistics versus $\bar{\mathrm{x}}_{HI}$. Notice that connected regions here show the average behaviour of all regions with ${x}_{HI} \gtrsim 0.5$ and holes refer to regions with  ${x}_{HI} \lesssim 0.5$ , from the way the limits of integration have been defined in eq.~\ref{eqn:rch} and \ref{eqn:betach}. From the best fit values shown in the bottom panel of table~\ref{table_nl}, we find that the average size of regions having $x_{HI} \lesssim0.5$ is smaller than the regions with $x_{HI} \gtrsim0.5$ while their number is less. Notice that this is when we are using the brightness temperature map which corresponds to $\bar{x}_{HI}=0.5$.  Since the maps have been smoothed at $R_s=9.5~\mathrm{Mpc}$, the values are not actual representation of the mean size as structures smaller than $R_s$ do not enter the averaging process. We also find that both connected regions and holes have $\beta_{hole} \simeq \beta_{con}$ at $\bar{x}_{HI}\simeq 0.5$.

	\begin{table}[H]
		\centering
		\small
		\begin{tabular}{||c|c|c|c||}
			\hline
			Parameter&Best fit&CV&Accuracy\\
			\hline
			\hline
			&&&\\
			$R_{mfp}~(\mathrm{Mpc})$&$18.93^{20.84}_{16.34}$&$0.16$&$0.05$\\
			
			&&&\\
			\hline
			&&&\\	
			$\zeta$&$18.57^{19.72}_{17.27}$&$0.066$&$0.06$\\
			
			&&&\\
			\hline
			&&&\\
			$T_{vir}~(\mathrm{K})$&$28797.20_{25347.70}^{31678.90}$&$0.11$&$0.04$\\
			
			&&&\\
			\hline
			&&&\\
			$\overline{x}_{HI}$&$0.467_{0.444}^{0.486}$&$0.047$&$0.066$\\
			&&&\\
			\hline
			\hline
		\end{tabular}
	\end{table}
	
	\begin{table}[H]
		\centering
		\begin{tabular}{||c|c||c|c||}
			\hline
			&&& \\
			Holes&Best fit&Connected& Best fit\\
			&&Regions& \\
			\hline
			\hline
			&&&\\
			$N_{hole}$&$155.18_{138.78}^{165.86}$&$N_{con}$&$120.27_{113.83}^{126.73}$\\
			
			&&&\\
			\hline
			&&&\\	
			$r^{ch}_{hole}$&$29.12_{27.11}^{31.72}$&$r^{ch}_{con}$&$34.40_{33.13}^{35.77}$\\
			
			&&&\\
			\hline
			&&&\\
			$\beta^{ch}_{hole}$&$0.56_{0.556}^{0.563}$&	$\beta^{ch}_{con}$&$0.57_{0.569}^{0.572}$\\
			
			&&&\\
			\hline
			&&&\\
			${A}^{ch}_{hole}$&$1885.75_{1700.14}^{2137.45}$&	$A^{ch}_{con}$&$2218.4_{2099.63}^{2348.99}$\\
			&&&\\
			\hline
			\hline
			
		\end{tabular}
		\caption{The table on top shows the best fit values for the model parameters of a noiseless mock observation smoothed at $R_s=9.5~\mathrm{Mpc}$. The best fit for $\bar{x}_{HI}$ is derived from these best fit parameters. The bottom table shows the corresponding derived values of the descriptors $N_{\mathrm{x}}$, $r^{ch}_{\mathrm{x}}$ and $\beta^{ch}_{\mathrm{x}}$. The upper and lower bounds show the $1-\sigma$ uncertainities. }
		\label{table_nl}
	\end{table}
	\begin{figure}
		\includegraphics[height=10.cm,width=14cm]{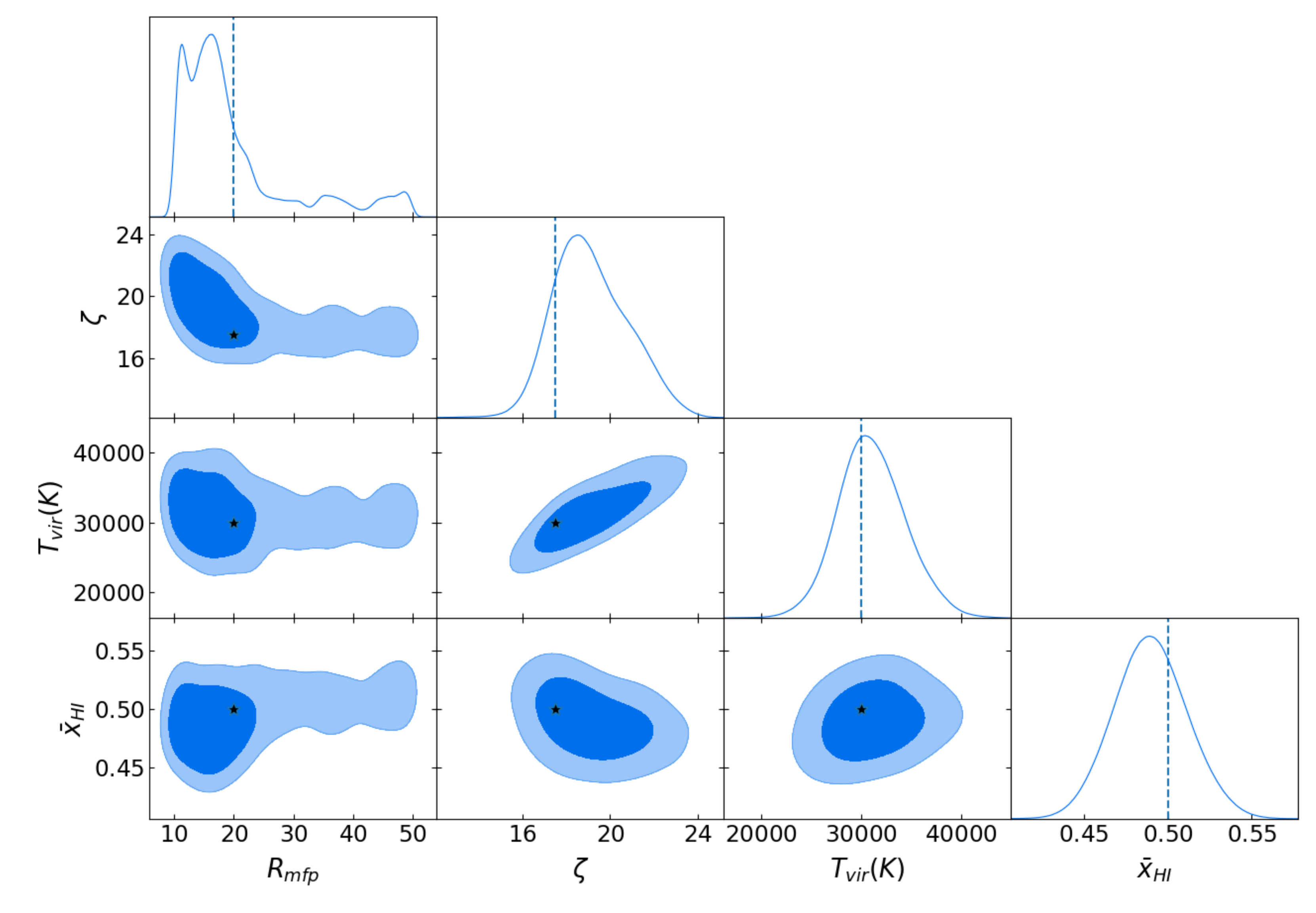} 
		
		\caption{Posterior probability distribution of parameters for a noisy mock $\delta T_b$ image at a redshift of $z=7.4$, smoothed at $R_s=9.5 \mathrm{Mpc}$ for an SKA observation with $t_{obs}=2000~\mathrm{hrs}$. The black star shows the input mock model ($\zeta=17.5, T_{vir}=3\times 10^4~\mathrm{K}$ and $R_{mfp}=20~\mathrm{Mpc}$). The value of $\bar{x}_{HI}=0.5$ is also marked. Note that the posterior distribution of $\bar x_{HI}$ is obtained by treating it as a derived parameter.} 
		\label{ell_noisy}
	\end{figure}
In fig.~\ref{ell_noisy} we show the triangle plots for the posterior probability distribution for a $\delta T_b$ map for an SKA noise for $t_{obs}=2000~\mathrm{hrs}$, smoothed at $R_s=9.5~\mathrm{Mpc}$. The best fit values for the same are shown in table~\ref{table_noisy}.  We also find that in order to recover our constraints to within $68\%$ credible intervalsfor a more feasible observation time of $t_{obs} < 2000~\mathrm{hrs}$, we will have to smooth our image to $R_s > 9.5~\mathrm{Mpc}$. Thus even a highly smoothed noisy map from SKA can give constraints to within permissible confidence interval. However, a high smoothing scale also implies loss of information at smaller scales and increased cosmic variance. Therefore, one would need to incorporate methods to mitigate noise, so as to recover even tighter constraints at lower smoothing scales and lesser $t_{obs}$.    
	
	

	\begin{table}
		\small
		
		\begin{tabular}{||c|c||}
			\hline
			Parameter&Best fit \\
			\hline
			\hline
			&\\
			$R_{mfp}~(\mathrm{Mpc})$&$19.87_{10.03}^{20.15}$\\
			
			&\\
			\hline
			&\\	
			$\zeta$&$19.13_{17.13}^{20.55}$\\
			
			&\\
			\hline
			&\\
			$T_{vir}~(\mathrm{K})$&$31091.7_{27545.6}^{34189.4}$\\
			
			&\\
			\hline
			&\\
			$\overline{x}_{HI}$&$0.489_{0.467}^{0.511}$\\
			&\\
			\hline
			\hline
			
		\end{tabular}
		\hspace{1.cm}
		\begin{tabular}{||c|c||c|c||}
			\hline
			&&& \\
			Holes&Best Fit&Connected& Best fit\\
			&&Regions& \\
			\hline
			\hline
			&&&\\
			$N_{hole}$&$173.75_{158.04}^{188.42}$&$N_{con}$&$118.21_{106.42}^{129.43}$\\
			
			&&&\\
			\hline
			&&&\\	
			$r^{ch}_{hole}$&$26.26_{23.87}^{28.26}$&$r^{ch}_{con}$&$34.37_{31.46}^{38.05}$\\
			
			&&&\\
			\hline
			&&&\\
			$\beta^{ch}_{hole}$&$0.56_{0.556}^{0.573}$&	$\beta^{ch}_{con}$&$0.568_{0.566}^{0.57}$\\
			
			&&&\\
			\hline
			&&&\\
			${A}^{ch}_{hole}$&$1612.51_{1352.63}^{1848.22}$&	$A^{ch}_{con}$&$2197.44_{1884.93}^{2583.37}$\\
			&&&\\
			\hline
			\hline
			
		\end{tabular}
		\caption{The left table shows the best fit and $1~\sigma$ bounds for the model parameters for a noisy mock $\delta T_b$ image at a redshift of $z=7.4$, smoothed at $R_s=9.5~\mathrm{Mpc}$ for an SKA observation with $t_{obs}=2000~\mathrm{hrs}$. The best fit value for $\bar{x}_{HI}$ is derived from these best fit parameters. The right table shows the corresponding derived values of the descriptors $N_{\mathrm{x}}$, $r^{ch}_{\mathrm{x}}$ and $\beta^{ch}_{\mathrm{x}}$. The upper and lower bounds show the $1-\sigma$ uncertainties.}
		\label{table_noisy}
	\end{table}
\section{Conclusion}\label{conc}
 Through this study we have described the prospects of constraining a basic EoR model using a combination of Minkowski tensors and Betti numbers. We performed our analysis on mock brightness temperature images from the Epoch of Reionization at $z=7.4$, with an ionization history such that $\bar{\mathrm{x}}_{HI}=0.5$ at that redshift. We find that the number count based statistic alone cannot be used to constrain model parameters due to degeneracy during different stages of ionization. However, when used in combination with morphological descriptors containing size and anisotropy information, we are able to obtain strong constraints on model parameters even at a high smoothing scale of $R_s=9.5~\mathrm{Mpc}$ . On using $\bar{\mathrm{x}}_{HI}$ as a derived parameter, we show that one can obtain tight constraints on the ionization history of the universe. We also showed that at $\bar{x}_{HI}=0.5$ the average size of regions with $x_{HI} \lesssim 0.5$ is smaller than regions with $x_{HI} \gtrsim 0.5$, while the regions are equally anisotropic. Finally, we showed that in order to recover our mock model to within $68 \%$ accuracy from an SKA-I low image at a single frequency channel of width $1~\mathrm{MHz}$, one would need to smooth the field at scales higher than $R_s=9.5~\mathrm{Mpc}$ in order to observe at $t_{obs}<2000~\mathrm{hrs}$ .

It is to be noted that the current analysis has been carried out on ideal images devoid of any foreground. We also performed our analysis at a single redshift and in a single frequency channel. The information content is expected to improve with the inclusion of more than one redshift with multiple frequency channels at each redshift. In a recent work \cite{sambit} the authors showed how using multiple frequency channels and performing the analysis using Betti numbers alone in 3D cubes can improve the observation time required for performing such an analysis. Moreover, we used averages of the statistics over field thresholds, thereby integrating the threshold information. The full potential of these descriptors can be exploited if we use the versus threshold information and perform the analysis by integrating our statistics in a 21cm based inference framework  \cite{gazagnes20} such as 21CMMC \cite{21cmmc}. The full analysis with inclusion of foreground and a wider astrophysical parameter space of EoR \cite {JP} will be carried out in a subsequent work in order to compare the constraints with those obtained from power spectrum in \cite{21cmmc}.  Our analysis using a combination of Minkowski tensors in 3D and Betti numbers on observed 3D cubes is underway. Through this study we have shown that a morphological description of even a single image at one redshift and a single frequency channel from the EoR can help us constrain EoR parameters and can serve as a promising statistical tool. 
\section*{Acknowledgment}
The computation required for this work was carried out on the the NOVA cluster at Indian Institute of Astrophysics and the BAEKDU cluster at the Center for Advanced Computing, Korea Institute for Advanced Study, Seoul, Republic of Korea. A.~Kapahtia and T.R. Choudhury acknowledge support of the Department of Atomic Energy, Government of India, under project no. 12-R\&D-TFR-5.02-0700. The work of P.~Chingangbam is supported by the Science and Engineering Research Board of the Department of Science and Technology, India, under the \texttt{MATRICS} scheme, bearing project reference no \texttt{MTR/2018/000896}. The work of R.~Ghara is supported by the Israel Science Foundation grant no. 255/18.  The work of S.~Appleby is supported by an appointment to the JRG Program at the APCTP through the Science and Technology Promotion Fund and Lottery Fund of the Korean Government, the Korean Local Governments in Gyeongsangbuk-do Province and Pohang City and by a KIAS Individual Grant QP055701 via the Quantum Universe Center at Korea Institute for Advanced Study, Seoul , Republic of Korea. We acknowledge use of the \texttt{GetDist} \cite{getdist} and \texttt{Matplotlib} \cite{matplotlib} packages. 


\begin{thebibliography}{99}
	
	\bibitem{loeb}
	A.~Loeb and S.~ R.~ Furlanetto:
	The First Galaxies in the Universe,
	Princeton Series in Astrophysics 
	(2013)
	
	\bibitem{Dayal:2018hft} 
	P.~Dayal and A.~Ferrara,
	Phys.\ Rept.\  {\bf 780-782}, 1 (2018)
	doi:10.1016/j.physrep.2018.10.002
	[arXiv:1809.09136 [astro-ph.GA]].
	\bibitem{gnedin} Gnedin, N.~Y.\ 2000, \ apj, 535, 530. doi:10.1086/308876
	\bibitem{Pritchard} Pritchard, J.~R. \& Loeb, A.\ 2012, Reports on Progress in Physics, 75, 086901. doi:10.1088/0034-4885/75/8/086901
	
	
	
	
	
	
	\bibitem{Bowman:2018} 
	Bowman, Judd D. ~Rogers, Alan E. E.
	~ Monsalve, Raul A.
	~Mozdzen, Thomas J.
	~ Mahesh, Nivedita
	''Nature {\bf 555}, 67 (2018)
	
	\bibitem{Parsons} 
	Parsons, A.~R., Backer, D.~C., Foster, G.~S., et al.\ 2010, aj, {\bf 139}, 1468
	\bibitem{Tingay et al.(2013)}
	Tingay, S.~J., Kaplan, D.~L., McKinley, B., et al.\ 2013, aj, {\bf 146}, 103
	\bibitem{van Haarlem}
	van Haarlem M.~P., et al., 2013, A\&A, {\bf 556}, A2
	\bibitem{Paciga:2011}
	Paciga, G. et al. , Mon. Not. Roy. Astron. Soc., {\bf 413}, 117 (2011)
	\bibitem{lofar} Mertens, F.~G., Mevius, M., Koopmans, L.~V.~E., et al.\ 2020, \ mnras, 493, 1662. doi:10.1093/mnras/staa327
	\bibitem{cooray} Cooray, A.\ 2005, \ mnras, 363, 1049. doi:10.1111/j.1365-2966.2005.09506.x
	\bibitem{Bharadwaj2005} Bharadwaj, S. \& Pandey, S.~K.\ 2005, \ mnras, 358, 968. doi:10.1111/j.1365-2966.2005.08836.x
    \bibitem{shaw}Shaw, A.~K., Bharadwaj, S., \& Mondal, R.\ 2019, \ mnras, 487, 4951. doi:10.1093/mnras/stz1561
	\bibitem{Banet} Banet, A., Barkana, R., Fialkov, A., et al.\ 2020, arXiv:2002.04956
	\bibitem{Shimabukuro} Shimabukuro, H., Yoshiura, S., Takahashi, K., et al.\ 2016, \ mnras, 458, 3003. doi:10.1093/mnras/stw48
	\bibitem{Watkinson_2017} Watkinson, C.~A., Majumdar, S., Pritchard, J.~R., et al.\ 2017, \ mnras, 472, 2436. doi:10.1093/mnras/stx2130
	
	\bibitem{suman}Majumdar, S., Pritchard, J.~R., Mondal, R., et al.\ 2018, \ mnras, 476, 4007. doi:10.1093/mnras/sty535
   \bibitem{Giri_bi} Giri, S.~K., D'Aloisio, A., Mellema, G., et al.\ 2019, \ jcap, 2019, 058. doi:10.1088/1475-7516/2019/02/058
	\bibitem{hutter2020} Hutter, A., Watkinson, C.~A., Seiler, J., et al.\ 2020, \ mnras, 492, 653. doi:10.1093/mnras/stz3139
	\bibitem{Gorce2019} Gorce, A. \& Pritchard, J.~R.\ 2019, \ mnras, 489, 1321. doi:10.1093/mnras/stz2195
		\bibitem{Maartens et al.(2015)}
		Maartens, R., Abdalla, F.~B., Jarvis, M., et al.\ 2015, Advancing Astrophysics with the Square Kilometre Array (AASKA14), 16,, arXiv e-prints , arXiv:1501.04076 
		\bibitem{Mellema} Mellema, G., Koopmans, L., Shukla, H., et al.\ 2015, Advancing Astrophysics with the Square Kilometre Array (AASKA14), 10
		\bibitem{giri18} Giri, S.~K., Mellema, G., Dixon, K.~L., et al.\ 2018, \ mnras, 473, 2949. doi:10.1093/mnras/stx2539
		\bibitem{giri19} Giri, S.~K., Mellema, G., Aldheimer, T., et al.\ 2019, \ mnras, 489, 1590. doi:10.1093/mnras/stz2224
		
		\bibitem{Kakiichi2017} Kakiichi, K., Majumdar, S., Mellema, G., et al.\ 2017, \ mnras, 471, 1936. doi:10.1093/mnras/stx1568
		\bibitem{Busch} Busch, P., Eide, M.~B., Ciardi, B., et al.\ 2020, \ mnras, 498, 4533. doi:10.1093/mnras/staa2599
		\bibitem{gazagnes20} Gazagnes, S., Koopmans, L.~V.~E., \& Wilkinson, M.~H.~F.\ 2020, arXiv:2011.08260
	\bibitem{Giri_super} Giri, S.~K., Mellema, G., \& Ghara, R.\ 2018, \ mnras, 479, 5596. doi:10.1093/mnras/sty1786
		
		\bibitem{Bag:2018zon} 
		S.~Bag, R.~Mondal, P.~Sarkar, S.~Bharadwaj and V.~Sahni,
		``The shape and size distribution of HII regions near the percolation transition,''
		Mon.\ Not.\ Roy.\ Astron.\ Soc.\  {\bf 477}, no. 2, 1984 (2018)
		doi:10.1093/mnras/sty714
		
		
		\bibitem{Bag:2018fyr} 
		S.~Bag, R.~Mondal, P.~Sarkar, S.~Bharadwaj, T.~R.~Choudhury and V.~Sahni,
		arXiv:1809.05520 [astro-ph.CO].
		
		
		\bibitem{Furlanetto:2016} Furlanetto, S.~R., \& Oh, S.~P.\ 2016, \ mnras, 457, 1813.
		
		\bibitem{Elbers} 
		W.~Elbers and R.~van de Weygaert,
		doi:10.1093/mnras/stz908
		arXiv:1812.00462 [astro-ph.CO].
		
		\bibitem{Lee:2007dt} 
		K.~G.~Lee, R.~Cen, J.~R.~Gott, III and H.~Trac,
		Astrophys.\ J.\  {\bf 675}, 8 (2008)
		[arXiv:0708.2431 [astro-ph]].
		
		\bibitem{Friedrich:2010nq} 
		M.~M.~Friedrich, G.~Mellema, M.~A.~Alvarez, P.~R.~Shapiro and I.~T.~Iliev,
		Mon.\ Not.\ Roy.\ Astron.\ Soc.\  {\bf 413}, 1353 (2011)
		
		\bibitem{Ahn:2010hg} 
		S.~E.~Hong, K.~Ahn, C.~Park, J.~Kim, I.~T.~Iliev and G.~Mellema,
		J.\ Korean Astron.\ Soc.\  {\bf 47}, no. 2, 49 (2014)
		[arXiv:1008.3914 [astro-ph.CO]].
		
		\bibitem{Wang:2015dna} 
		Y.~Wang, C.~Park, Y.~Xu, X.~Chen and J.~Kim,
		Astrophys.\ J.\  {\bf 814}, no. 1, 6 (2015)
		[arXiv:1510.01404 [astro-ph.CO]].
		
		\bibitem{Gleser:2006su} 
		L.~Gleser, A.~Nusser, B.~Ciardi and V.~Desjacques,
		Mon.\ Not.\ Roy.\ Astron.\ Soc.\  {\bf 370}, 1329 (2006)
		doi:10.1111/j.1365-2966.2006.10556.x
		[astro-ph/0602616]. 
		
		\bibitem{Yoshiura:2015} S. Yoshiura, H. Shimabukuro, K. Takahashi and T. Matsubara, arxiv:1602.0235.
		
		
		\bibitem{Chen} 
		Z.~Chen, Y.~Xu, Y.~Wang and X.~Chen,
		arXiv:1812.10333 [astro-ph.CO].
	


	
	
	
	
	\bibitem{Schroder2D:2009}
	G.E. Schroder-Turk, S. Kapfer, B. Breidenbach, C. Beisbart, and K. Mecke,
	{{\em J. Microsc.} {\bf 238} 57 (2010)}.
	
	
	\bibitem{Vidhya:2016} V.~Ganesan and P. Chingangbam, 
	{{\em JCAP} {\bf 1706} 023 (2017)} 
	
	\bibitem{Chingangbam:2017} P.~Chingangbam, K.~P.~Yogendran, Joby P.~K.~, V.~Ganesan and Stephen Appleby, Changbom Park, 
	{{\em JCAP} {\bf 1712} 023 (2017)}
	
	
	\bibitem{Appleby:2017uvb} 
	S.~Appleby, P.~Chingangbam, C.~Park, S.~E.~Hong, J.~Kim and V.~Ganesan,
	Astrophys.\ J.\  {\bf 858}, no. 2, 87 (2018)
	doi:10.3847/1538-4357/aabb53
	[arXiv:1712.07466 [astro-ph.CO]].
	\bibitem{Goyal} Goyal, P., Chingangbam, P., \& Appleby, S.\ 2020, \ jcap, 2020, 020. doi:10.1088/1475-7516/2020/02/020
	
	
	
	
	
	
	
	
	
	
	
	\bibitem{Kapahtia:2017qrg} 
	A.~Kapahtia, P.~Chingangbam, S.~Appleby and C.~Park,
	JCAP {\bf 1810}, no. 10, 011 (2018)
	doi:10.1088/1475-7516/2018/10/011
	[arXiv:1712.09195 [astro-ph.CO]].
	
	\bibitem{Kapahtia2019} Kapahtia, A., Chingangbam, P., \& Appleby, S.\ 2019, \ jcap, 2019, 053. doi:10.1088/1475-7516/2019/09/053

	
	
	
	\bibitem{Mesinger:2010ne} 
	A.~Mesinger, S.~Furlanetto and R.~Cen,
	Mon.\ Not.\ Roy.\ Astron.\ Soc.\  {\bf 411}, 955 (2011)
	[arXiv:1003.3878 [astro-ph.CO]].
	\bibitem{cosmomc} Lewis, A. \& Bridle, S.\ 2002, \ prd, 66, 103511. doi:10.1103/PhysRevD.66.103511
	
	
		
		\bibitem{ghara2017} Ghara, R., Choudhury, T.~R., Datta, K.~K., et al.\ 2017, \ mnras, 464, 2234. doi:10.1093/mnras/stw2494
	\bibitem{Tomita:1986} H.~Tomita, Progr.~Theor.~Phys.~{\bf 76}, 952 (1986)
	\bibitem{Schmalzing:1997}
	J. Schmalzing and T. Buchert, 
	{{\em Astrophys. J.} {\bf 482} L1-L4 (1997)}
	
	
	
\bibitem{Schmalzing:1998}
	J. Schmalzing and K. M. Gorski, 
	\bibitem{Field(1958)}
	Field, G.~B.\ 1958,
	Proceedings of the IRE, 46, 240
	\bibitem{Raghu_sim} Ghara, R., Choudhury, T.~R., \& Datta, K.~K.\ 2015, \ mnras, 447, 1806. doi:10.1093/mnras/stu2512
	\bibitem{miralda} Chen, X. \& Miralda-Escud{\'e}, J.\ 2004, \ apj, 602, 1. doi:10.1086/380829
	\bibitem{xray} Ross, H.~E., Dixon, K.~L., Ghara, R., et al.\ 2019, \ mnras, 487, 1101. doi:10.1093/mnras/stz1220 
	\bibitem{lya}Ghara, R. \& Mellema, G.\ 2020, \ mnras, 492, 634. doi:10.1093/mnras/stz3513
	\bibitem{Furlanetto:2006jb} 
	S.~Furlanetto, S.~P.~Oh and F.~Briggs,
	``Cosmology at Low Frequencies: The 21 cm Transition and the High-Redshift Universe,''
	Phys.\ Rept.\  {\bf 433}, 181 (2006)

		\bibitem{Zeldovich:1969sb} 
		Y.~B.~Zeldovich,
		Astron.\ Astrophys.\  {\bf 5}, 84 (1970).
	
	
	
	\bibitem{Furlanetto:2004nh} 
	S.~Furlanetto, M.~Zaldarriaga and L.~Hernquist,
	Astrophys.\ J.\  {\bf 613}, 1 (2004)
	doi:10.1086/423025
	[astro-ph/0403697]
	\bibitem{Sheth}{2001MNRAS.323....1S} Sheth, R.~K., Mo, H.~J., \& Tormen, G.\ 2001, \ mnras, 323, 1. doi:10.1046/j.1365-8711.2001.04006.x

	\bibitem{Thompson} Richard Thompson, A., Moran, J. M.,  Swenson Jr, G. W. (2017). Interferometry and synthesis in radio astronomy (p. 872). Springer Nature.
	
	

	
	
	\bibitem{Furlanetto:2006tf} 
	S.~Furlanetto,
	Mon.\ Not.\ Roy.\ Astron.\ Soc.\  {\bf 371}, 867 (2006)
	[astro-ph/0604040].

		\bibitem{JP} Park, J., Mesinger, A., Greig, B., et al.\ 2019, \ mnras, 484, 933. doi:10.1093/mnras/stz032
	\bibitem{21cmmc} Greig, B. \& Mesinger, A.\ 2017, \ mnras, 472, 2651. doi:10.1093/mnras/stx2118
\bibitem{getdist} Lewis, A.\ 2019, arXiv:1910.13970
\bibitem{matplotlib} J. D. Hunter, "Matplotlib: A 2D Graphics Environment," in Computing in Science \& Engineering, vol. 9, no. 3, pp. 90-95, May-June 2007, doi: 10.1109/MCSE.2007.55.
\bibitem{sambit} Giri, S.~K. \& Mellema, G.\ 2020, arXiv:2012.12908
\end{thebibliography}


%



\end{document}